\documentclass[aps,prc,numerical,showpacs,preprint]{revtex4-1}
\usepackage{amsmath, amssymb, graphics, setspace}%
\usepackage[mathscr]{eucal}%
\usepackage{epsfig}
\usepackage{epstopdf} %
\usepackage{subfigure}%
\usepackage{dcolumn}%
\usepackage{multirow}%
\newcommand{\mathsym}[1]{{}}
\newcommand{\unicode}[1]{{}}
\begin{document}
	\title{Temperature effects on nuclear pseudospin symmetry in the Dirac-Hartree-Bogoliubov formalism}
	\author{R. Lisboa}
	\affiliation{Universidade Federal do Rio Grande do Norte, \\
		Escola de Ci\^encias e Tecnologia, \\
		59078-970 Natal, Rio Grande do Norte, Brazil}
	\email{ronai@ect.ufrn.br}
	\author{P. Alberto}
	\affiliation{CFisUC, University of Coimbra, Physics Department,\\
		P-3004-516 Coimbra, Portugal}
	\author{B. V. Carlson}
	\affiliation{Departamento de F\'isica, \\
		Instituto Tecnol\'ogico da Aeron\'autica, \\
		Centro T\'ecnico Aeroespacial,
		12228-900 S\~ao Jos\'e dos Campos,\\
		S\~ao Paulo, Brazil}
	\author{M. Malheiro}
	\affiliation{Departamento de F\'isica, \\
		Instituto Tecnol\'ogico da Aeron\'autica, \\
		Centro T\'ecnico Aeroespacial,
		12228-900 S\~ao Jos\'e dos Campos,\\
		S\~ao Paulo, Brazil}
	\date{\today}
\begin{abstract}
We present finite temperature Dirac-Hartree-Bogoliubov (FTDHB) calculations 
for the tin isotope chain to study the dependence of pseudospin on 
the nuclear temperature.  In the FTDHB calculation, the density dependence of the 
self-consistent relativistic mean fields, the pairing, and the vapor phase 
that takes into account the unbound nucleon states are considered 
self-consistently. The mean field potentials obtained in the FTDHB 
calculations are fit by Woods-Saxon (WS) potentials to examine how the 
WS parameters are related to the energy splitting of the pseudospin pairs as 
the temperature increases. We find that the nuclear potential surface 
diffuseness is the main driver 
for the pseudospin splittings and that it increases  as the
temperature grows. We conclude that pseudospin symmetry is better realized 
when the nuclear temperature increases. 
The results confirm the findings of previous works using relativistic mean field theory 
at $T=0$, namely 
that the correlation between the pseudospin splitting and the parameters of the 
Woods-Saxon potentials implies that pseudospin symmetry is a dynamical
symmetry in nuclei. We show that the dynamical nature of the pseudospin 
symmetry remains when the temperature is considered in a realistic calculation 
of the tin isotopes, such as that of the Dirac-Hartree-Bogoliubov formalism.
\end{abstract}
\maketitle
%
\section{Introduction}
\label{sec:Intro}

Since the seminal article published by Ginocchio,
\cite{Gino97}, pseudospin symmetry has been extensively studied 
in relativistic mean field (RMF) and relativistic Hartree-Fock (RHF) 
theories, with the intention of understanding the origin
of pseudospin symmetry and its symmetry breaking.

The evidence for pseudospin symmetry comes from nuclear energy spectra with
quasi degeneracy between pairs of single-particle states with quantum numbers
$(n, l, j = l+1/2)$ and $(n-1,l+2,j=l+3/2)$ in a spherical 
basis where, $n$, $l$,
and $j$ are the radial, orbital, and total angular momentum
quantum numbers,
respectively, of the upper component of the Dirac spinor. Pseudospin symmetry
was recognized as a relativistic symmetry when Ginocchio point out the
pseudospin doublets can be written as
$(\tilde{n}=n-1,\tilde{l}=l+1,\tilde{j}=\tilde{l}\pm 1/2)$ where,
the quantum numbers $\tilde{n}$, $\tilde{l}$, and $\tilde{j}$ are the
quantum numbers of the lower component of the Dirac spinor
\cite{Gino97,Gino99}. Pseudospin symmetry is exact when the doublets
with $j=\tilde{l}\pm\tilde{s}$ are degenerate.

In RMF theory the Dirac equation with 
attractive scalar, $V_S(r)$,
and repulsive vector, $V_V(r)$, potentials displays exact
pseudospin symmetry when
$\Sigma(r) = V_S(r) + V_V (r) = 0$ or more generally, when
$\Sigma^{\prime}(r)=d\Sigma(r)/dr=0$ \cite{Meng98,Liang16}.
For finite nuclei, the $\Sigma(r)$ field plays the role of 
the nuclear binding potential in a relativistic theory. 
Thus, the bound states cannot exist for $\Sigma(r)=0$
when we are considering spherically symmetric potentials that 
vanish at large distances \cite{Gino97}.
However, exact pseudospin symmetry is possible when $\Delta(r)=V_V(r)-V_S(r)$ is a
spherical relativistic harmonic oscillator potential because 
it does not tend to zero at large distances \cite{Meng03,Lisboa04,Gino04}.
Recently, it has been shown that this behavior is shared by general 
radial potentials that tend to infinity at large distances \cite{Alberto13}.
In this case, $\Delta(r)$ acts as a binding potential
in the second-order differential equation of the 
lower component of Dirac spinor, because it acts as an effective mass that goes to infinity.
Thus, in this case the pseudospin symmetry is exact and there are 
still bound states.

For realistic nuclei with nuclear mean fields which 
vanish at large distances,
the cancellation between scalar and vector potentials gives a relatively 
small binding potential of about $\Sigma\approx -60$ MeV at the center and 
thus, pseudospin symmetry cannot be exact.
The purpose the studies performed in most
works on pseudospin symmetry is to understand
its origin and its symmetry-breaking
\cite{Savushkin06,Liang15}.

The Dirac equation has been solved for different potentials and systems to
study how the pseudospin doublets become degenerate or almost degenerate.
Usually, the pseudospin splitting depends on the shape of the 
potentials that are used to solve the Dirac equation.
In previous works \cite{Alberto01, Alberto02}, the Woods-Saxon 
potential was used because the conditions $\Sigma(r) = 0$ 
and $\Sigma^{\prime}(r)=0$ can be
met approximately by varying the parameters of this potential.
The pseudospin splitting depends on the depth of $|\Sigma_0|$,
its surface diffuseness and its radius. Then, the authors
reduced the Dirac equation into two Schr\"odinger-like equations for the lower
(and upper) spinor component, each
being a sum of different terms: kinetic, pseudospin-orbit (and spin-), a
Darwin term, and potential terms with $\Sigma(r)$ and $\Delta(r)$ potentials.
By taking the expectation values of these terms, one obtains 
an energy decomposition
for each single-particle level, allowing the study of the 
non-perturbative nature
of pseudospin symmetry. The pseudospin-orbit term has the denominator
$E-\Sigma(r)$ and thus becomes infinite when $E=\Sigma(r)$, but the singularity
is canceled by kinetic and $\Sigma$ terms originating in the
quasi-degeneracy \cite{Alberto02}.
In fact, a cancellation exists between the large pseudospin-orbit potential
and other terms, showing the dynamic character of the pseudospin 
symmetry and its
non-perturbative nature \cite{Savushkin06,Liang15,Alberto01,Alberto02}.

  Before this understanding of the non-perturbative nature of the
pseudospin symmetry, the condition $\Sigma^{\prime}(r)=0$, which appears
in the pseudospin-orbital term, was interpreted as favoring the
restoration of pseudospin symmetry due its competition with the
centrifugal barrier \cite{Meng99}. For exotic nuclei with highly
diffuse potentials, $\Sigma^{\prime}(r)\approx 0$ may be
a good approximation and then the pseudospin symmetry will be
good \cite{Meng06}. In this case, to study exotic nuclei, it is
necessary to use a
relativistic continuum Hartree Bogoliubov (RCHB)
theory that properly considers the pairing correlations and the coupling
to the continuum via the Bogoliubov transformation in a microscopic and
self-consistent way \cite{Meng06}. This approach is also useful when studying
exotic nuclei with unusual $N/Z$ ratios, where the neutron 
(or proton) Fermi surface
is close to the particle continuum. The contribution
of the continuum and/or resonances is then important \cite{Vretenar05}. In
Ref.~\cite{Zhang04}, the pseudospin symmetry of the 
resonant states in $^{208}$Pb 
was calculated by solving the Dirac equation with Woods-Saxon-like
vector and scalar potentials using the coupling-constant method.
It was found that the diffusivity of the potentials plays a significant
role in the energy splitting and the width of the resonant pseudospin partners.
In Ref.~\cite{Lu12}, the pseudospin symmetry in single-particle resonant states 
in nuclei was also shown to be exactly conserved under the
same condition as for the pseudospin symmetry in bound 
states, i.e., $\Sigma(r)=0$ or $\Sigma^{\prime}(r)=0$. 

It is well accepted that RCHB theory can be used to study
pairing correlations due to the short-range part of the
nucleon-nucleon interaction in open shell nuclei, as well as to describe the
exotic nuclei. However, the calculations of finite nuclei can be better
performed when the pairing correlations, the nucleon, and mesons
mean fields are all calculated self-consistently and this is not done for
the pairing field in RCHB calculations, where the pairing correlation
is introduced in a non-relativistic way as a Skyrme-type $\delta$ force or
finite-range Gogny force \cite{Meng06}.  
In Ref.~\cite{Carlson00} the 
self-consistent Dirac-Hartree-Bogoliubov (DHB) approach was introduced to 
self-consistently include pairing energy and gaps in 
calculations for spherical and 
deformed nuclei. 
As an extension, we have applied the DHB approach to hot nuclei
including finite-temperature effects to study spherical and deformed nuclei
and to analyze how the binding energy, the neutron and charge radii,
the deformation and, in particular the pairing gap change with
temperature \cite{Lisboa07}. We introduce in our finite temperature DHB
(FTDHB) calculation a vapor subtraction procedure to take into account the 
contribution of the resonant nucleon states and to remove long-range Coulomb 
repulsion between the hot nucleus and the gas as well as the 
contribution of the 
external nucleon gas \cite{Bonche84,Bonche85,Lisboa10}. Quite recently, we 
found that for small temperatures the vapor subtraction procedure is not very 
relevant to the change of the pairing fields with increasing 
temperature because the
critical superfluidity and superconducting phase transitions occur at
$T\sim 1$ MeV. The effects of the vapor phase that takes into account the 
unbound nucleon states become important only at temperatures $T\geq 4$ MeV, 
allowing the study of nuclear properties of finite nuclei from zero
to high temperatures \cite{Lisboa16}.

As in RCHB theory, the advantage of FTDHB to study pseudospin 
symmetry is that the
particle levels for the bound states in the canonical basis are the same as
those coming from solving the Dirac equation with scalar and vector
potentials from RMF \cite{Meng06}. The form of the radial 
equations for the lower
and upper components of the Dirac equation remain the same in 
the canonical basis
even after the pairing interaction has been taken into account \cite{Meng06}.
Furthermore, another advantage of FTDHB calculations lies in the fact that it 
considers the proper isospin dependence of the spin-orbit term, 
as well as the isospin 
and energy dependence of the pseudospin symmetry \cite{Meng06}.
In non-selfconsistent RMF calculations, the
isospin asymmetry of the nuclear pseudospin comes mainly from the
vector-isovector $V_{\rho}$ potential and its effect on different terms
of the Schr\"odinger-like equation contributing to the pseudospin 
splittings cancel 
each other to a certain extent \cite{Lisboa03}.
In Ref.~\cite{Long640}, a density-dependent RHF (DDRHF) theory 
for nuclear systems  was introduced without dropping the Fock terms. 
Thus, the coupling was taken to be a function of the baryonic density, 
as well as considering the pion--nucleon coupling, which is effective only 
through exchange terms. 
The contributions of the $\sigma$, $\omega$, and $\rho$mesons in 
this DDRHF are much smaller than their corresponding ones in RMF 
\cite{Long640}. The Fock terms change the effective mass that 
contains the scalar part of the nucleon self-energy as well the 
vector potential. These effects could play some role on the symmetry 
of pseudospin, which depends on $\Sigma(r)=V_S(r)+V_V(r)$. 
The same authors show that the Fock terms bring 
significant contributions to the pseudospin orbital potential, 
but these contributions are canceled by other exchange terms due to 
the non-locality of the exchange potentials \cite{Long639}. 
As a result, the pseudospin symmetry is preserved even considering 
the Fock terms.
On the other hand, the density dependence of the 
self-consistent relativistic mean fields and pairing fields, as well 
as the vapor phase that considers the unbound nucleon states, allows us 
to analyze in a realistic way the effect of temperature on the quasi 
degeneracy of pseudospin partners.

In this work, we use FTDHB calculations to study the temperature dependence
of mean-field potentials and its effects on pseudospin symmetry. The
attractive scalar, $V_S(r)$, and repulsive vector $V_V(r)$, potentials obtained
in our calculations have a shape very similar to a Woods-Saxon one. We fit
the central potential $\Sigma_c$ mean field, as well as the total potential
for neutrons and protons with a Woods-Saxon shape for each tin isotope in 
order to better assess how temperature changes the Woods-Saxon parameters: 
the depth of potential, the radius $R$, and the surface diffuseness $a$. 
In RMF theory at temperature zero, 
there is a correlation 
between the pseudospin-orbit term and the pseudospin energy splitting when the 
radius, diffusivity, and the depth of potential are 
varied \cite{Alberto01, Alberto02, Lisboa03}.  
We will show that the magnitude of the pseudospin doublets splitting 
decreases with increasing temperature and that the behavior of the parameters 
of the Woods-Saxon potential for $T\neq 0$ obeys the same systematics as 
for $T = 0$.

 We use the tin nuclei as a function of the number of nucleons
from $A = 100$ to $170$ and temperatures varying from $T=0$ up to $T=8$ MeV.
These tin isotopes allow us to apply our study to the stable and
unstable nuclei from proton drip line to the neutron drip line.
The pseudospin symmetry was investigated before in these exotic nuclei 
using a RCHB calculation but at $T=0$ \cite{Meng99}. For $T\neq 0$, these
nuclei were also used to study the evolution of the pairing gaps
and critical temperature along isotopic and isotonic chains of
semi magic nuclei in FTDHB \cite{Lisboa16} and FTRHFB \cite{Li2015} calculations.

The paper is organized as follows. In Sec. \ref{sec:two} we present briefly the 
formalism of the finite temperature Dirac-Hartree-Bogoliubov 
model. In Sec. \ref{sec:results} 
we present and discuss the results of the calculations and in 
Sec. \ref{sec:conclusion} we draw our conclusions.
\section{The formalism}
\label{sec:two}
We use the self-consistent Dirac-Hartree-Bogoliubov (DHB) formalism of
Ref.~\cite{Carlson00}, but we consider explicitly the self-consistent 
temperature dependence of the relativistic pairing fields, as well 
as the vapor phase,
to take into account the unbound nucleon states. This finite-temperature DHB
(FTDHB) formalism was developed in an earlier work \cite{Lisboa16} and includes
the Coulomb and meson mean fields, as well as pairing correlations, 
to calculate
the properties of hot nuclei self-consistently. 
As discussed before, the Fock terms are neglected, although the most important effects 
of the Fock terms, due to exchange of the short-range $\sigma$,
$\omega$, and $\rho$  mesons, can be taken into account by using 
adjusted Hartree terms.
The Hamiltonian form is given by
\begin{equation}\label{eq:hamloc}
\left(\begin{array}{cc}
\varepsilon+\mu_{t}-h_{t}(\vec{x}) & \bar{\Delta}_{t}^{\dagger}(\vec{x})\\
\bar{\Delta}_{t}(\vec{x}) & \varepsilon-\mu_{t}+h_{t}(\vec{x})
\end{array}\right)
\left(
\begin{array}{c}
\mathcal{U}_{t}(\vec{x})\\
\gamma_{0}\mathcal{V}_{t}(\vec{x})
\end{array}\right)=0\,,\qquad\qquad t=p,n,
\end{equation}
where, in the diagonal terms, $\varepsilon$ denotes the quasi-particle
energies, $\mu_t$ represents the chemical potential to be used as a 
Lagrange multiplier to fix the average number of protons $(t=p)$ and
neutrons $(t=n)$, and $h_t$ stands for the single-particle Hamiltonian of the
nucleon. The
nondiagonal terms, $\overline{\Delta}_t$ and its conjugate 
$\overline{\Delta}_t^{\dagger}$, are the pairing fields, which account for
correlated pairs of time-reversed single-particle states, i.e,
the paired particle-particle states. 
The components $\mathcal{U}_t(\vec{x})$ and $\mathcal{V}_t(\vec{x})$
represent the Dirac spinors corresponding to the normal and 
time-reversed components, respectively. We write each of the four-component
spinors as
\begin{equation} \label{eq:DSGFu}
\mathcal{U}_{t\alpha }(\vec{x})=
\left(
\begin{array}{l}
   G_{\mathcal{U},t\alpha}(\vec{x}) \\
i\,F_{\mathcal{U},t\alpha}(\vec{x})
\end{array}
\right)\,,
\quad \mbox{and}\quad
\gamma _{0}\mathcal{V}_{t\alpha}(\vec{x})=
\left(
\begin{array}{l}
   G_{\mathcal{V},t\alpha}(\vec{x}) \\
i\,F_{\mathcal{V},t\alpha}(\vec{x})
\end{array}
\right)\,.
\end{equation}
The Dirac Hamiltonian is
\begin{equation}
h_{t}(\vec{x})=-i\vec{\alpha}\cdot\vec{\nabla}+\beta M^{*}(\vec{x})+V_{t}(\vec{x})\,,
\end{equation}
where the effective mass $M^{*}$ contains the scalar part of the nucleon 
self-energy from the Dirac field and $V_{t}$ is the vector potential. 
These are written as
\begin{eqnarray}\label{eq:vANDs}
M^{*}(\vec{x})&=&M-g_{\sigma}\,\sigma(\vec{x})\,\\
V_{t}(\vec{x})&=&g_{\omega}\,\omega^{0}(\vec{x})+\frac{g_{\rho}}{2}\,2m_{t}\,\rho^{00}(\vec{x})+e\,
\left(\frac{1}{2}+m_{t}\right)\, A^{0}(\vec{x})\,.
\end{eqnarray}
The constant $M$ is the nucleon mass, while
$g_{\sigma}$, $g_{\omega}$, $g_{\rho}$, and $e$ are the corresponding
coupling constants for the mesons and the photon. The isospin projections
are $m_t=1/2$ for protons and $m_t=-1/2$ for neutrons. The fields $\omega^0$
and  $A^0$ are the timelike components of the four-vector $\omega$ and
photon fields, while $\rho^{00}$ is the third component of 
the timelike component of the isovector-vector $\rho$ meson,
\begin{eqnarray}
\omega^{0}(\vec{x}) & = & g_{\omega}\int d^{3}z\, d_{\omega}^{0}(\vec{x}-\vec{z})\rho_{B}(\vec{z})\,,\nonumber \\
\rho^{00}(\vec{x}) & = & \frac{g_{\rho}}{2}\int d^{3}z\, d_{\rho}^{0}(\vec{x}-\vec{z})\rho_{3}(\vec{z})\,,\nonumber \\
A^{0}(\vec{x}) & = & e\int d^{3}z\, d_{\gamma}^{0}(\vec{x}-\vec{z})\rho_{c}(\vec{z})\,,\\
\sigma(\vec{x}) & = & g_{\sigma}\int d^{3}z\, d_{\sigma}(\vec{x}-\vec{z})\rho_{s}(\vec{z})\nonumber \\
& = & \int d^{3}z\, d_{\sigma}^{0}(\vec{x}-\vec{z})\left(g_{\sigma}\rho_{s}(\vec{z})-g_{3}\,\sigma(\vec{x})^{2}-g_{4}\,
\sigma(\vec{x})^{3}\right)\,.\nonumber
\end{eqnarray}
where the propagators are
\begin{equation}
d_{j}^{0}(\vec{x}-\vec{z})=
\frac{1}{4\pi\,\left|\vec{x}-\vec{z}\right|}\times
\left\{
\begin{array}{cl}
1  									&,\quad\mbox{for photons}\\
\exp\left(-m_{j}\left|\vec{x}-\vec{z}\right|\right)     &,\quad\mbox{for mesons.}
\end{array}
\right.\label{eq:prop0}
\end{equation}
The Hartree contributions to the self-energy can be written 
in terms of the normal densities,
\begin{eqnarray}\label{eq:rhost}
\rho_{s}(\vec{x},T) & = &
 2\sum_{\varepsilon_{t\alpha}<0,t}
 \left(
  \mathcal{U}_{t\alpha}^{\dagger}\gamma_{0}\mathcal{U}_{t\alpha}n( \varepsilon_{t\alpha},T)
 +\mathcal{V}_{t\alpha}^{\dagger}\gamma_{0}\mathcal{V}_{t\alpha}n(-\varepsilon_{t\alpha},T)
 \right),\nonumber \\
\rho_{B}(\vec{x},T) & = & 
 2\sum_{\varepsilon_{t\alpha}<0,t}
\left(
 \mathcal{U}_{t\alpha}^{\dagger}\mathcal{U}_{t\alpha}n( \varepsilon_{t\alpha},T)
+\mathcal{V}_{t\alpha}^{\dagger}\mathcal{V}_{t\alpha}n(-\varepsilon_{t\alpha},T)
\right),\nonumber \\
\rho_{3}(\vec{x},T) & = & 
2\sum_{\varepsilon_{t\alpha}<0,t}2m_{t}\,
\left(
 \mathcal{U}_{t\alpha}^{\dagger}\mathcal{U}_{t\alpha}n( \varepsilon_{t\alpha},T)
+\mathcal{V}_{t\alpha}^{\dagger}\mathcal{V}_{t\alpha}n(-\varepsilon_{t\alpha},T)
\right),\nonumber \\
\rho_{c}(\vec{x},T) & = & 
2\sum_{\varepsilon_{t\alpha}<0,t}(m_{t}+1/2)\,
\left(
 \mathcal{U}_{t\alpha}^{\dagger}\mathcal{U}_{t\alpha}n( \varepsilon_{t\alpha},T)
+\mathcal{V}_{t\alpha}^{\dagger}\mathcal{V}_{t\alpha}n(-\varepsilon_{t\alpha},T)
\right)\;.
\end{eqnarray}
     The Hamiltonian form of the pairing field is,
\begin{eqnarray}\label{eq:pardbefore}
\bar{\Delta}_{t}^{\dagger}(\vec{x})&=&
\gamma_{0}\Delta_{t}(\vec{x})\gamma_{0}\nonumber \\
&=&c_{pair}
\left(
\frac{g_{\sigma}^{2}}{m_{\sigma}^{2}}\gamma_{0}\,\kappa_{t}(\vec{x},T)\,\gamma_{0}
\right.\nonumber\\
&-&
\left.
\left(
\frac{g_{\omega}^{2}}{m_{\omega}^{2}}+\frac{(g_{\rho}/2)^{2}}{m_{\rho}^{2}}
\right)
\gamma_{0}\gamma^{\mu}\,\kappa_{t}(\vec{x},T)\,\gamma_{\mu}\gamma_{0}
\right)\,.
\label{eq:deloc}
\end{eqnarray}
where we neglect its Coulomb and nonlinear $\sigma$-meson contributions. We
approximate the contributions of the other mesons using the zero-range
limit of the meson propagators. The zero-range approximation greatly simplifies 
the numerical calculations, but must be calibrated phenomenologically.
Thus, an overall constant $c_{pair}$ has been introduced in the expression 
for the pairing field to compensate for deficiencies of the interaction parameters 
and of the numerical calculation \cite{Carlson00}. 
In Ref.~\cite{Lisboa16}, we emphasize that this is not a weakness of our 
calculations alone, but of any Hartree-(Fock)-Bogoliubov calculation using a 
limited space of states and an effective interaction, even those using a 
finite-range one. We studied the deficiencies of RMF 
meson-exchange interactions for the description of pairing in nuclear matter 
and nuclei in detail in Ref. \cite{Carlson1997}, 
where we examined the strong correlation between the position of 
the $NN$ virtual state in the vacuum and the magnitude of the pairing field 
in nuclear matter. Our conclusion was that RMF interactions 
do not describe pairing correctly because they do not describe low-energy 
$NN$ scattering correctly. The use of RMF interactions in the 
particle-particle channel can be corrected by introducing a cutoff,  
multiplying by a constant factor, or both \cite{Carlson00}. 
Furthermore, to take into account correctly 
the density-dependent competition between scalar and vector interactions 
that occurs in both the mean field and the pairing, we use 
a fully relativistic interaction in the pairing channel \cite{Carlson2003,Carlson2007}.

The anomalous density $\kappa_{t}(\vec{x},T)$ is given by
\begin{eqnarray}
\kappa_{t}(\vec{x},T) &=&
\frac{1}{2}\sum_{\varepsilon_{t\gamma}<0}
\left(
\mathcal{U}_{t\gamma}(\vec{x})\overline{\mathcal{V}}_{t\gamma}(\vec{x})
+\gamma_{0}B\mathcal{V}_{t\gamma}^{*}(\vec{x})\mathcal{U}_{t\gamma}^{T}(\vec{x})B^{\dagger}
\right)\,\nonumber\\
&&\times
\left(n(\varepsilon_{t\gamma},T)-n(-\varepsilon_{t\gamma},T)\right),\label{eq:anomadens}
\end{eqnarray}
where $B=\gamma_5\,C$ and $C$ is the charge conjugation matrix that provides the
time-reversed Dirac structure of the wave vectors.
 
 For both normal and anomalous densities, one sees that the temperature
enters in our calculation only through the Fermi occupation factors
\begin{equation}
n(\varepsilon_{\gamma},T)=\frac{1}{1+\exp\left(\varepsilon_{\gamma}/T\right)}\;,\label{eq:fermiocc}
\end{equation}
where $\varepsilon_{\gamma}$ represent the quasiparticle energy.
Thus, the temperature dependence of a solution of the FTDHB
equation comes from the quasi-particle normal and anomalous densities.
When $T\rightarrow 0$, the Fermi occupation factors are
$n(\varepsilon_{\gamma},T)=1$ and $n(-\varepsilon_{\gamma},T)=0$ and we
recover the usual nuclear densities of a finite nucleus. 
The quasiparticle energies that enter each Fermi occupation factor 
have opposite signs. Thus,
as $T$ increases, there is a reduction of the anomalous density due to the
difference between the two contributions to the Fermi occupation factor, as we
see in Eq.~(\ref{eq:anomadens}). As a consequence, the pairing energy and gap
tend to zero as the temperature increases \cite{Lisboa16}.

Specifically, for axially symmetric potentials, the scalar and vector
potential are independent of the azimuthal angle such that 
$V_{S,V}=V_{S,V}(r_{\bot},z)$. 
We have that $V_{S,V}(r_{\bot},z)\rightarrow 0$ for $r_{\bot}\rightarrow \infty$ 
or $z\rightarrow \pm\infty$ and $r_{\bot}V_{S,V}(r_{\bot},z)\rightarrow 0$ 
for $r_{\bot}\rightarrow 0$ \cite{Leviatan11}. Furthermore,
the rotational symmetry is broken when we 
chose this axial symmetry, but the densities are invariant with respect 
to a rotation around the symmetry axis. As a consequence the projection of 
the total angular momentum along the symmetry axis $\Omega_{\alpha}$, as well as 
the parity $\pi$ and the isospin projection $t$, are still good quantum numbers. 
Because of to this and the time-reversed Dirac structure, the two equal and 
opposite values of angular momentum projection $\pm\Omega_{\alpha}$ are 
degenerate in energy.

 The Dirac spinors in Eqs.~(\ref{eq:DSGFu}) take the forms 
\begin{equation} \label{eq:DSGFu2}
\mathcal{U}_{t\alpha }(\vec{x})
=\frac{1}{\sqrt{2\pi }}\left(
\begin{array}{l}
   G_{\mathcal{U},t\alpha }^{+}(r_{\bot},z)\,e^{i(\Omega _{\alpha }-1/2)\varphi } \\
   G_{\mathcal{U},t\alpha }^{-}(r_{\bot},z)\,e^{i(\Omega _{\alpha }+1/2)\varphi } \\
i\,F_{\mathcal{U},t\alpha }^{+}(r_{\bot},z)\,e^{i(\Omega _{\alpha }-1/2)\varphi } \\
i\,F_{\mathcal{U},t\alpha }^{-}(r_{\bot},z)\,e^{i(\Omega _{\alpha }+1/2)\varphi }
\end{array}
\right) 
\,
\end{equation}
and
\begin{equation} \label{eq:DSGFv}
\gamma _{0}\mathcal{V}_{t\alpha}(\vec{x})
=\frac{1}{\sqrt{2\pi }}\left(
\begin{array}{l}
   G_{\mathcal{V},t\alpha }^{+}(r_{\bot},z)\,e^{i(\Omega _{\alpha }-1/2)\varphi } \\
   G_{\mathcal{V},t\alpha }^{-}(r_{\bot},z)\,e^{i(\Omega _{\alpha }+1/2)\varphi } \\
i\,F_{\mathcal{V},t\alpha }^{+}(r_{\bot},z)\,e^{i(\Omega _{\alpha }-1/2)\varphi } \\
i\,F_{\mathcal{V},t\alpha }^{-}(r_{\bot},z)\,e^{i(\Omega _{\alpha }+1/2)\varphi }
\end{array}
\right) 
\,.
\end{equation}

Thus, the radial wave functions $G_{\mathcal{U,V}}^{\pm}(r_{\bot},z)$ 
and $F_{\mathcal{U,V}}^{\pm}(r_{\bot},z)$ and the meson fields are 
expanded in terms of the eigenfunctions of a deformed axially symmetric 
harmonic oscillator:
\begin{equation}
V_{osc}(z,r_{\bot})=\frac{1}{2}M(\omega^2_z z^2 +\omega_{\bot}^2r_{\bot}^2)
\end{equation}
where the oscillator frequencies $\hbar\omega_{z}$ and
$\hbar\omega_{\bot}$ are written in terms of a deformation 
parameter $\beta_0$, as
\begin{equation}
\hbar\omega_z = \hbar\omega_0 e^{-\sqrt{5/(4\pi)}\beta_0}\quad\mbox{and}\quad
\hbar\omega_{\bot} = \hbar\omega_0 e^{+\frac{1}{2}\sqrt{5/(4\pi)}\beta_0}\,.
\end{equation}
The $(z,r_{\bot})$ dependence of 
eigenfunctions in large and small components of the Dirac spinors are 
divided by oscillator length,
\begin{equation}
b_z = \sqrt{\hbar/M\omega_z}\quad\mbox{and}\quad
b_{\bot} = \sqrt{\hbar/M\omega_{\bot}}\,,
\end{equation}
and because of volume conservation it is guaranteed $b_z b_{\bot}^2=b_0^3$.
The parameter $b_0=\sqrt{\hbar/M\omega_0}$ stands for the oscillator length
corresponding to the oscillator frequency $\hbar\omega_0$ of the
spherical case. In this way, the spherical and deformed basis are
determined by oscillator frequency $\hbar\omega_0$ and deformation $\beta_0$.
Thus, the method can be applied to both spherical and axially deformed
nuclei. 

Inserting these expansions of eigenfunctions into the
Dirac-Gorkov equation (\ref{eq:hamloc}), we can reduce the equation to 
the diagonalization problem of a symme\-tric matrix and calculate the 
Hartree densities of Eq.~(\ref{eq:rhost}) and the components of the 
anomalous density of Eq.~(\ref{eq:pardbefore}). 
The fields of the massive mesons are obtained by solving the Klein-Gordon 
equations using a similar expansion with the same deformation parameter 
$\beta_{0}$ but a smaller oscillator length of $b_{B}=b_{0}/\sqrt{2}$.
The Coulomb field is calculated directly in configuration space.

This method is a direct generalization of the one described 
in Refs.~\cite{Carlson00,Lisboa16,Gambir90,Lalazissis98,Vretenar98,Lalazissis99} 
where more details can be found.
%
\section{Results}
\label{sec:results}

In this section, we present FTDHB calculations for hot nuclei to investigate 
the effect of temperature in the mean field potentials and its 
consequences for the pseudospin symmetry. To study the effect of 
temperature in the mean field potentials, we examine the tin isotopes 
from $A=100$ to $A=170$. 
We use the nonlinear Walecka model with the 
NL3 interaction because it allows us a comparison
with our calculation at temperature zero \cite{Lalazissis96}.  
The study of nuclear systems has been made recently using 
new types of parametrizations such as density-dependent meson-nucleon couplings 
(DD-ME1) \cite{Niksic02} as well as point-coupling interaction (PC-PK1) 
\cite{Zhao10}.
These interactions have been 
used to study, for instance, the paring interaction at finite temperature 
\cite{Li2015}. The results are consistent  
with those obtained by us at temperatures of about $1 - 2$ MeV, where the
pairing interaction is important, 
and also for hot nuclei at temperatures above 
2 MeV \cite{Lisboa16,Gambhir00,Calrson04}.
In our calculations,
the expansion of harmonic oscillator basis is truncated at a finite number of
major shells, with the quantum number of the last included shell set by
$N_F=14$ in the case of the fermions and by $N_B = 24$ for the bosons. These
bases are sufficient to achieve convergence in our numerical calculation
and reproduce experimental and earlier theoretical results of
the literature at both low and high temperatures. In all cases, the oscillator
frequencies $\hbar\omega_0=\hbar\omega_z=\hbar\omega_{\bot}=41A^{-1/3}$ MeV,
corresponding to an undeformed basis, were used. A value of the overall
constant $c_{pair}=0.55$ was introduced in the pairing interaction 
for neutrons and protons, Eq.~(\ref{eq:pardbefore}), which due to the 
self-consistency, results
in a null pairing field, as expected for the closed-shell nuclei we are
studying. This means we are studying the spherical tin isotopes 
from $A=100$ to $A=170$. 
Among them, the nuclei $^{100}$Sn, $^{132}$Sn, and $^{176}$Sn have pairing gap
and energy zero, so that pairing has no effect on pseudospin symmetry over 
the entire range of temperatures considered.  For open-shell and 
deformed nuclei, the nuclear pairing energy and gap vanishes above the 
relatively  low temperatures of $T=0.5 - 1.2$ MeV \cite{Li2015,Lisboa16}.

\begin{figure*}[!t]
	\subfigure{\label{fig:VxASnN}
		\includegraphics[width=7.5cm,height=7.5cm]{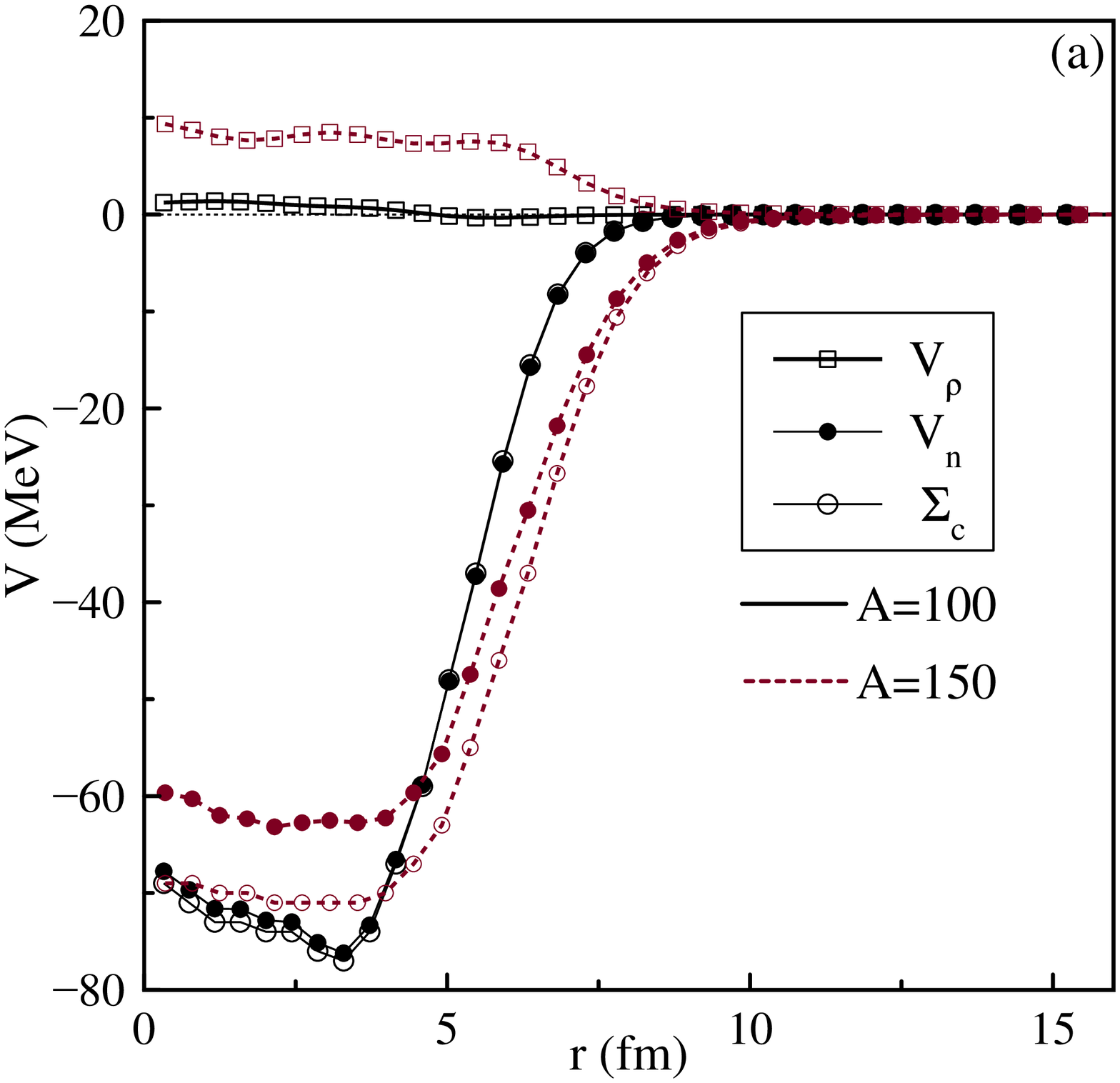}}\hfill
	\subfigure{\label{fig:VxASnP}
		\includegraphics[width=7.5cm,height=7.5cm]{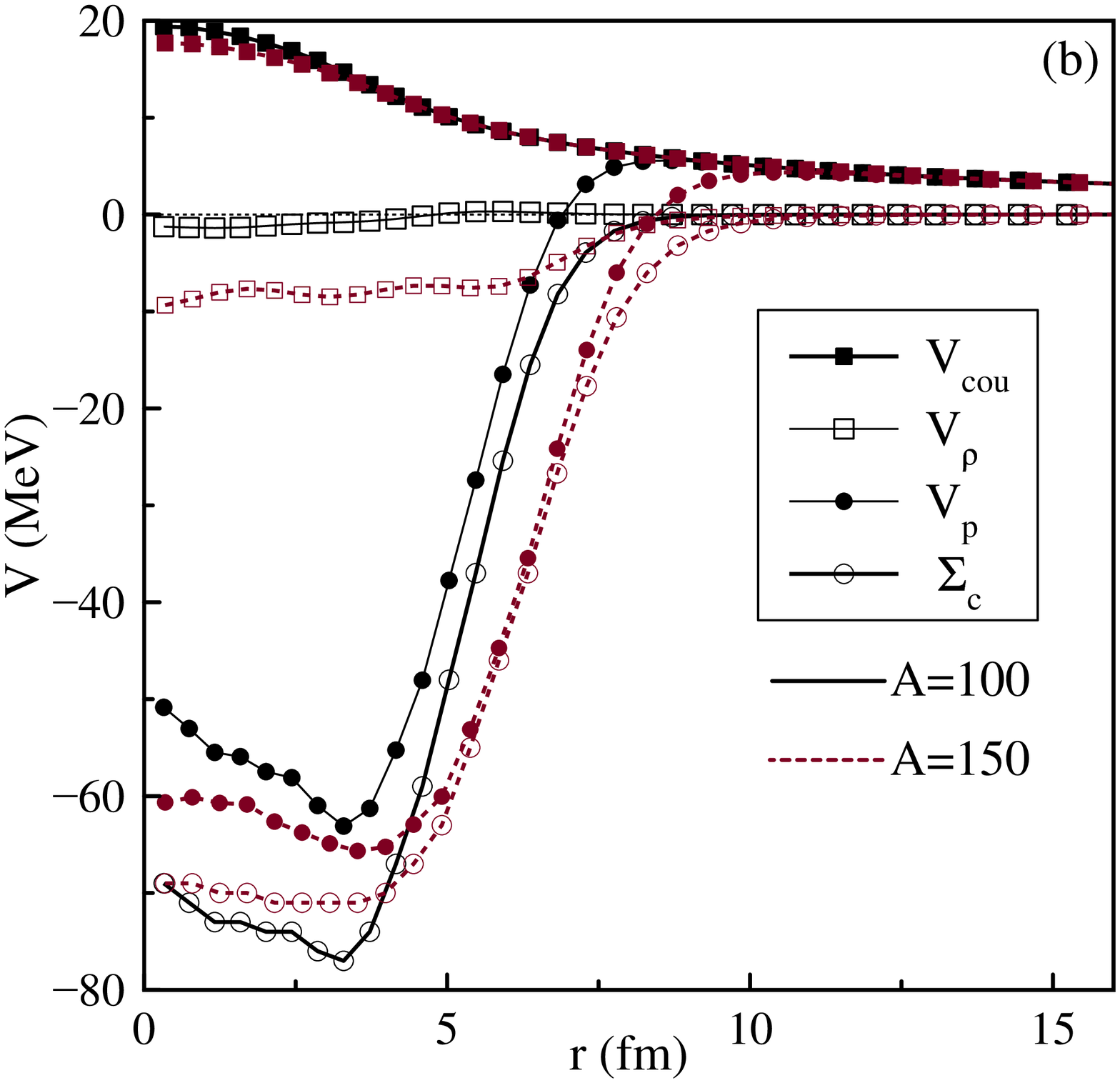}}\hfill
	\subfigure{\label{fig:VnxASn}
		\includegraphics[width=7.5cm,height=7.5cm]{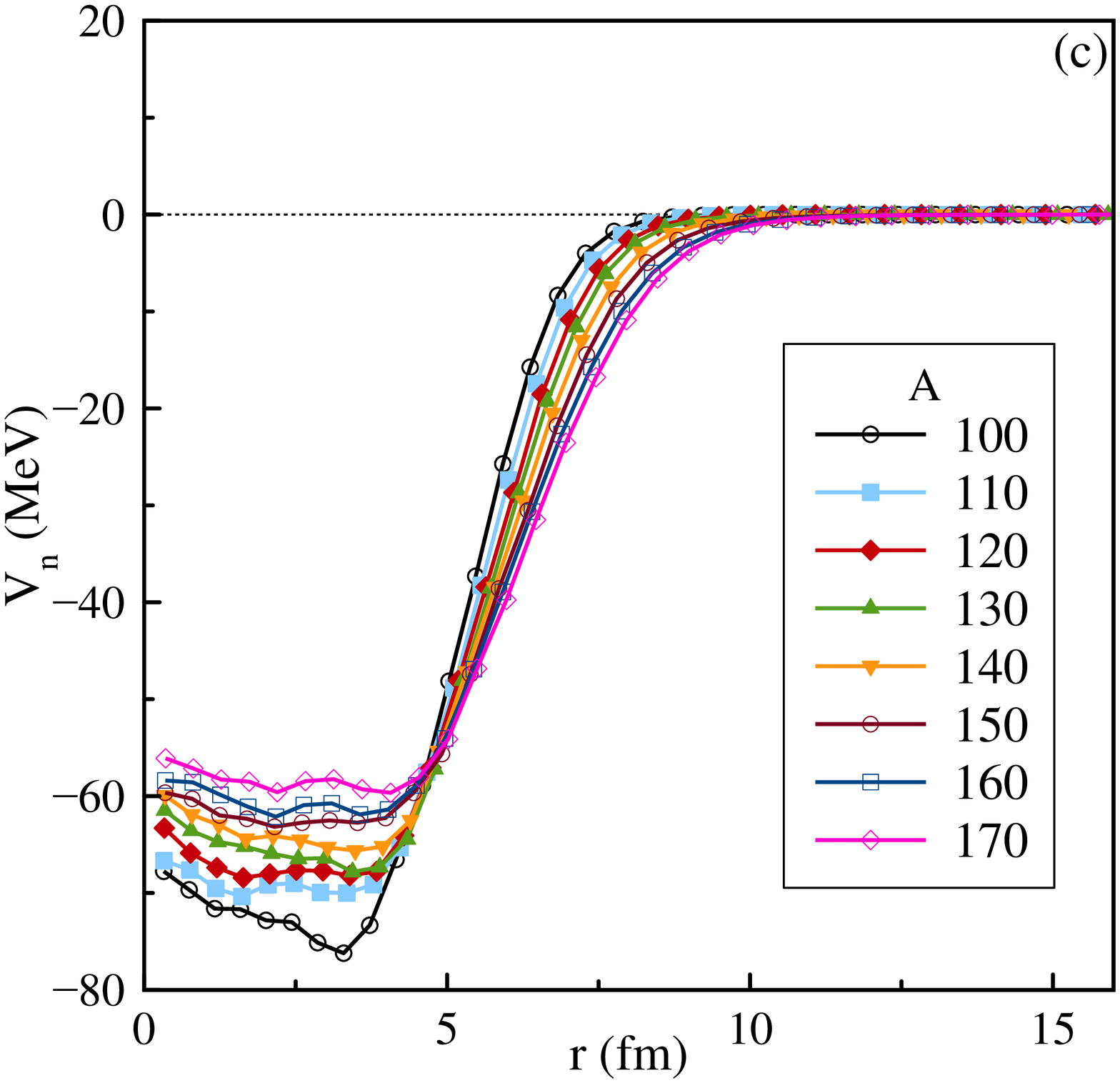}}\hfill
	\subfigure{\label{fig:VpxASn}
		\includegraphics[width=7.5cm,height=7.5cm]{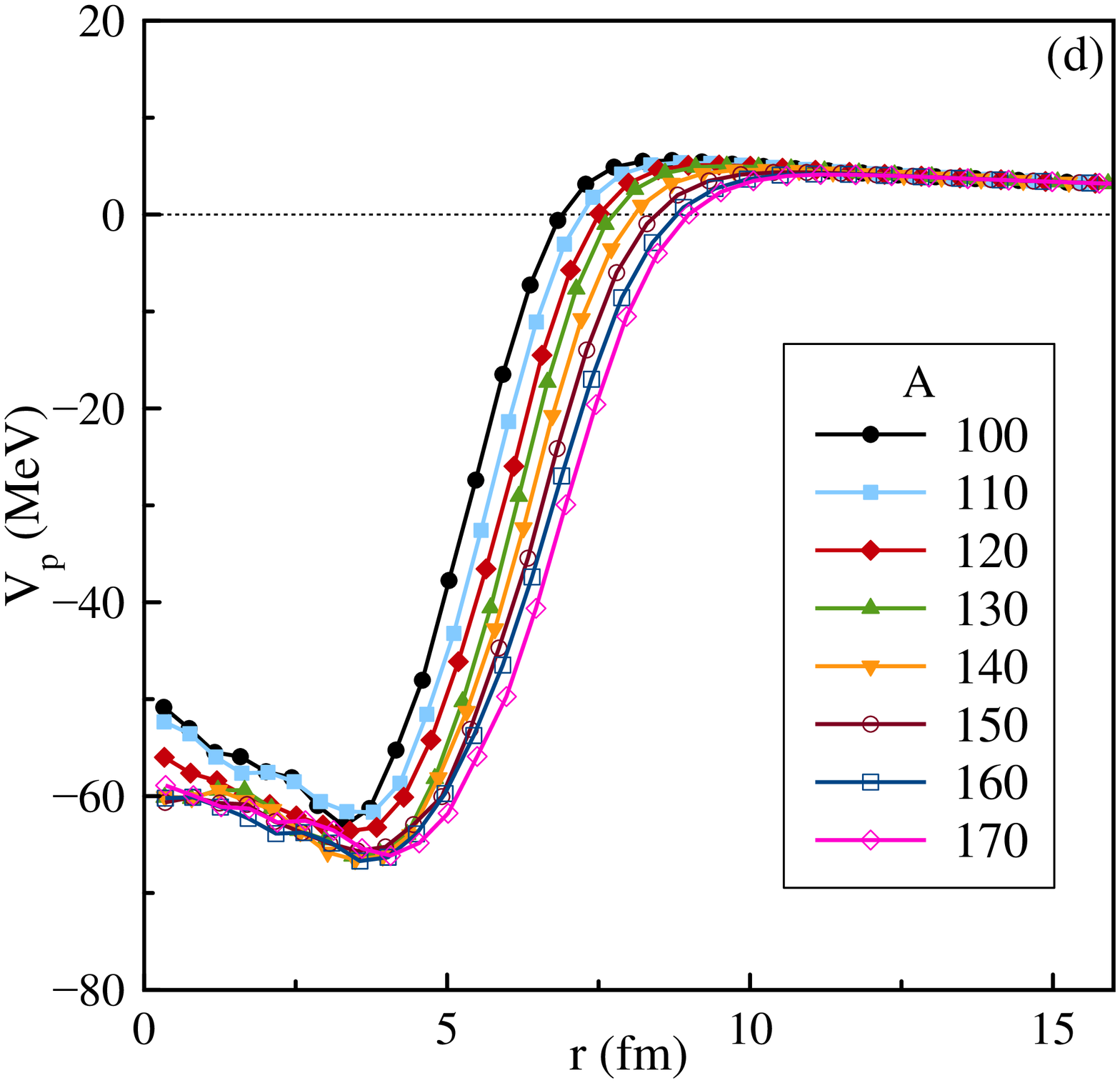}}\hfill
	\caption{\label{fig:VSn100} (Color online) Nuclear potentials 
as a function of the radial distance for the tin isotope chain. In the 
top panels, the meson potentials for (a) neutrons and (b) protons with 
the  Coulomb potential are displayed for $^{100}$Sn (full lines) 
and $^{150}$Sn (dashed lines). In the bottom panels, the potentials 
(c) $V_n(r)$ for neutrons and (d) $V_p(r)$ for protons are shown 
from $A=100$ to $A=170$ . \vspace{-0.5cm}}
\end{figure*}

In Fig.~\ref{fig:VxASnN}, we show the potentials 
$V_{\rho}(r)$, $\Sigma_{c}(r)=V_{\sigma}(r)+V_{\omega}(r)$, 
and $V_n(r)=\Sigma_{c}(r)-V_{\rho}(r)$,
as a function of the radial distance for $^{100}$Sn and $^{150}$Sn at $T=0$. 
The full lines represent the nucleus $^{100}$Sn, for which we see 
that the $V_{\rho}(r)$ potential (empty squares) is very small, while its sum 
with $\Sigma_{c}(r)$ (empty circles) produces a shallow 
potential $V_n(r)$ (full circles) for neutrons. The same behavior can 
be seen for $^{150}$Sn,
represented by the dashed lines, but now $V_{\rho}(r)$ is large and 
as a consequence, $V_n(r)$ is more affected by it.
In Fig.~\ref{fig:VxASnP}, the
$V_{\rho}(r)$ potential has the opposite sign, the repulsive Coulomb potential
$V_{coul}(r)$ has a long range and their sum shifts
the potential $V_p(r)=\Sigma_{c}(r)+V_{\rho}(r)+V_{coul}(r)$ for protons. 
Because of the small
$V_{\rho}(r)$ potential in comparison to $V_{coul}(r)$ for $^{100}$Sn, the large
difference between $V_{p}(r)$ and $\Sigma_{c}(r)$ at the nuclear center is due 
practically to $V_{coul}(r)$ alone.
For $^{150}$Sn, the contribution of $V_{\rho}(r)$ is significant 
in comparison to
$V_{coul}(r)$, and as a consequence there is a cancellation between the two 
that produces a difference of the same order of magnitude between $V_{p}(r)$ 
and $\Sigma_c(r)$.
In the DHB calculation,
the nuclear potentials for protons and neutrons
for the case $N=Z$ ($^{100}$Sn) are not the same. The Coulomb potential
changes the proton energy levels and, because of that, in the 
self-consistent DHB calculation the neutron energy levels are also 
changed in such a way that there is a net $V_{\rho}$ potential \cite{Lisboa03}.
%

In Fig.~\ref{fig:VnxASn}, we show the potential $V_n(r)$ for 
neutrons, and in Fig.~\ref{fig:VpxASn},
the potential $V_{p}$(r) for protons as a function of
radial distance for the tin isotope chain from $A=100$ to $A=170$. These
results, obtained in a FTDHB self-consistent calculation, 
show that the mean-field
potentials have the shape of a Woods-Saxon potential.
In Refs.~\cite{Alberto01,Alberto02}, RMF studies at $T=0$ were performed 
to investigate
the correlation between the pseudospin splitting and the parameters
of the Wood-Saxon potential: its depth ($\Sigma_0$), surface diffuseness ($a$),
and radius ($R$). In Ref. \cite{Lisboa03} this was done for a 
single isotope chain.  The neutron $V_n(r)$ and proton $V_p(r)$ 
mean-field potentials in a
tin isotope chain were parameterized by a Woods-Saxon form as functions of $A$.
For the tin isotopes as $A$ increases, the central potential $|\Sigma_0|$
decreases and the surface diffuseness increases, effects which both favor
the pseudospin symmetry. However, the radius increases with $A$, 
which can partially
offset those effects \cite{Alberto01}.  Since the values $|\Sigma_0|R^2$ 
are roughly constant for neutrons, the correlation between these two values,
mentioned above, implies that the effects of increasing $R$ and
decreasing $|\Sigma_0|$ in the neutron central potential, when A increases,
balance each other. Thus, the dominant effect comes from the increasing
value of $a$, slightly favoring the pseudospin 
symmetry \cite{Lisboa03}. However,
for protons, the value $|\Sigma_0|R^2$ is not constant, because both
$|\Sigma_0|$ and the radius $R$ increase as $A$ increases for the tin isotopes
Hence, the changes in $|\Sigma_0|$ and $R$ disfavor the
pseudospin symmetry in this case. The isospin asymmetry in pseudospin
symmetry is due to the isovector $V_{\rho}$ potential, which is repulsive
for neutrons and attractive for protons and makes the vector potential $V_V$
bigger for neutrons than for protons. As a consequence, $|\Sigma_0|$ becomes
smaller for neutrons than for protons \cite{Alberto01,Lisboa03}.

\begin{figure*}[!ht]
	\subfigure{\label{fig:VcxTSn100} 			
		\includegraphics[width=6.3cm,height=6.3cm]{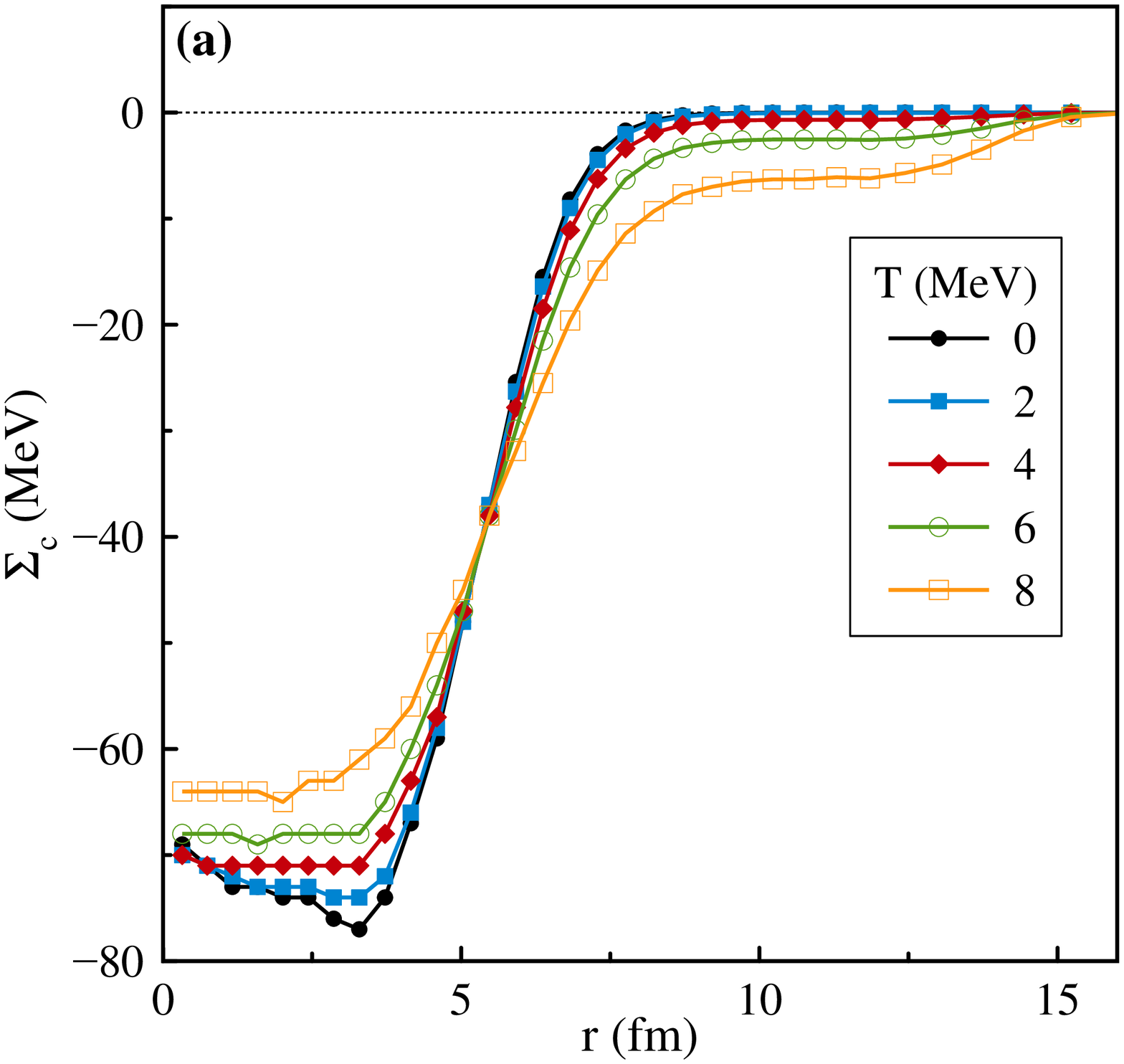}}\hfill
	\subfigure{\label{fig:VcxTSn100FIT}
		\includegraphics[width=6.3cm,height=6.3cm]{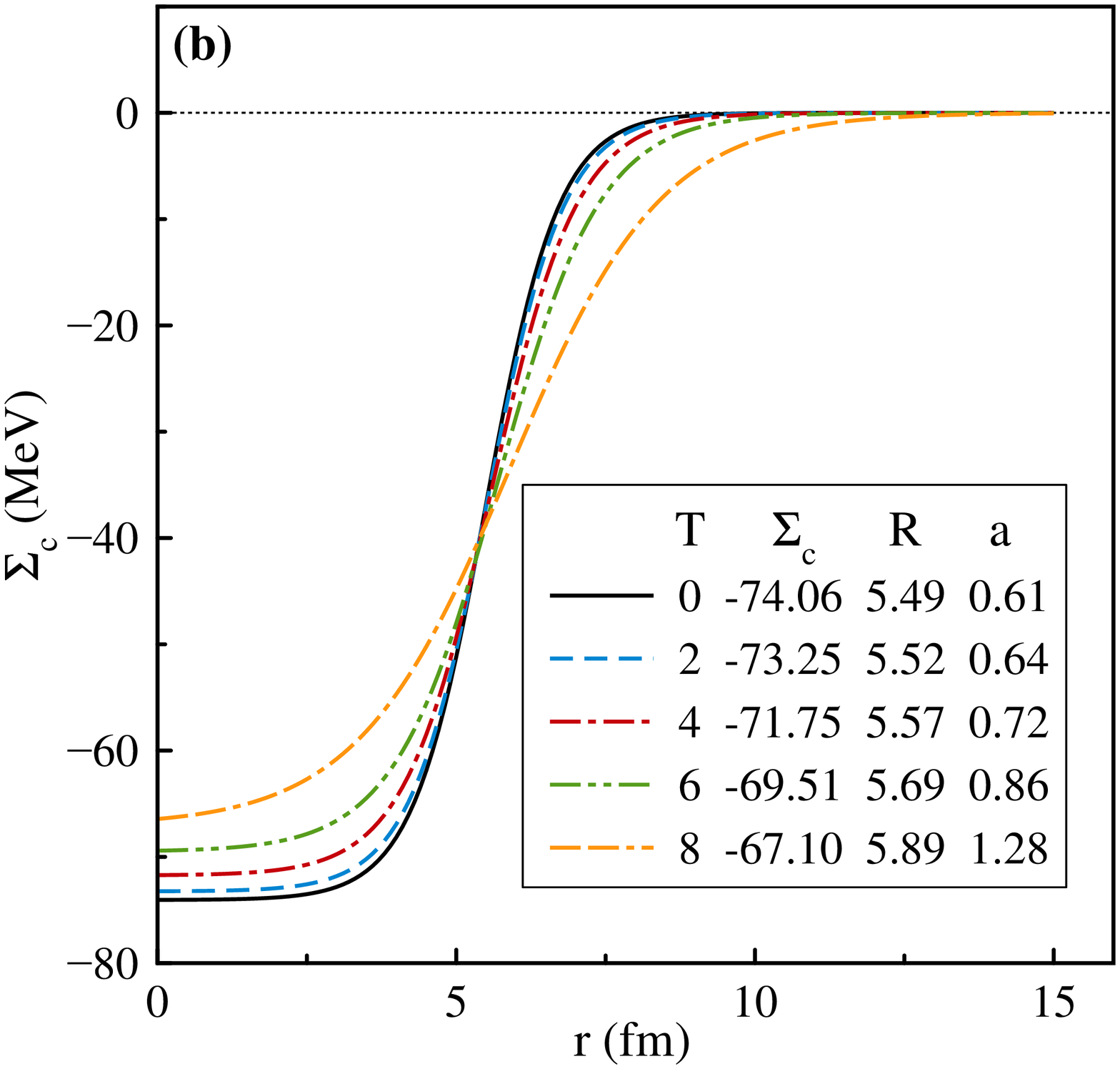}}\hfill
	\subfigure{\label{fig:VnxTSn100}
		\includegraphics[width=6.3cm,height=6.3cm]{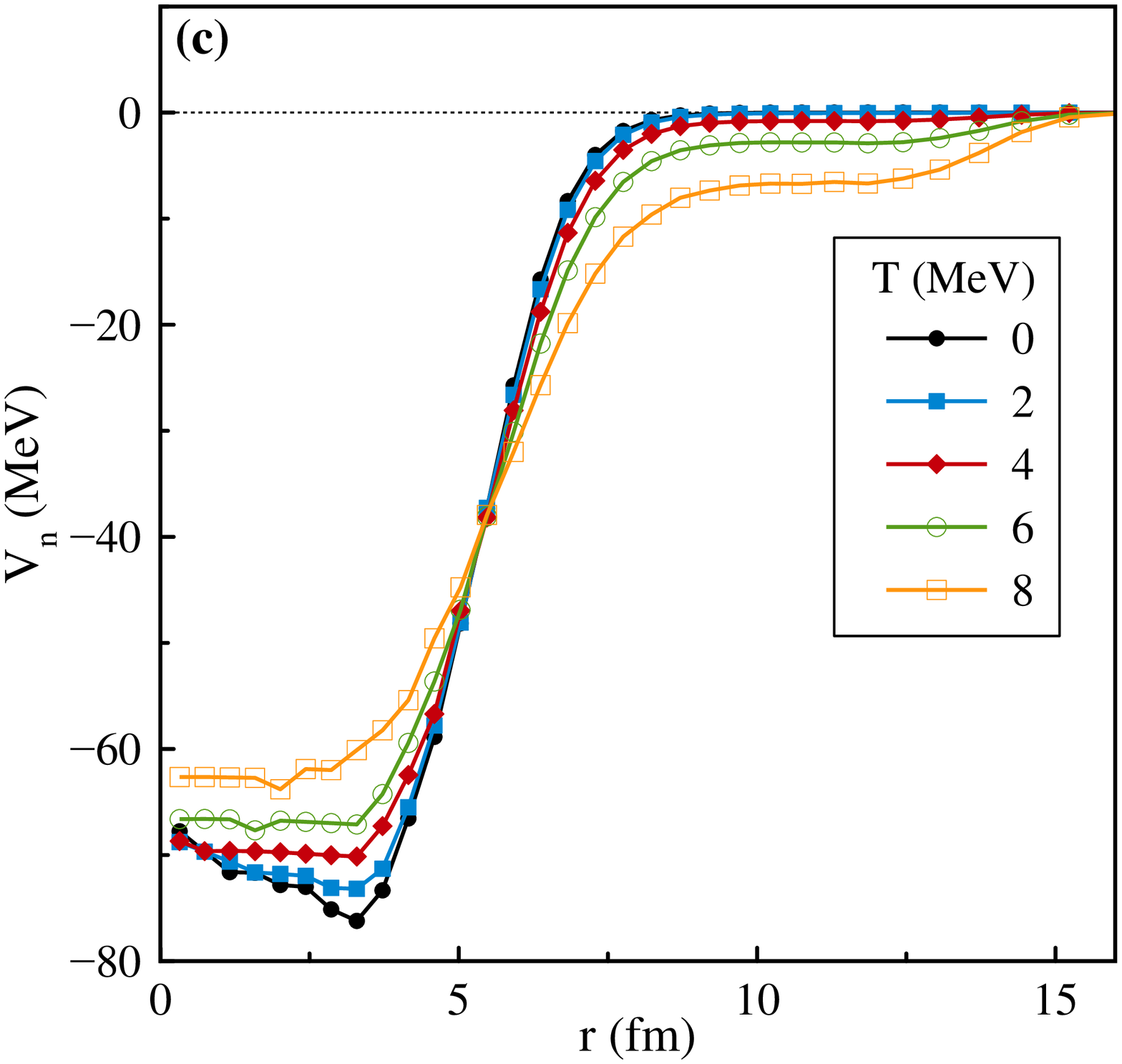}}\hfill
	\subfigure{\label{fig:VnxTSn100FIT}
		\includegraphics[width=6.3cm,height=6.3cm]{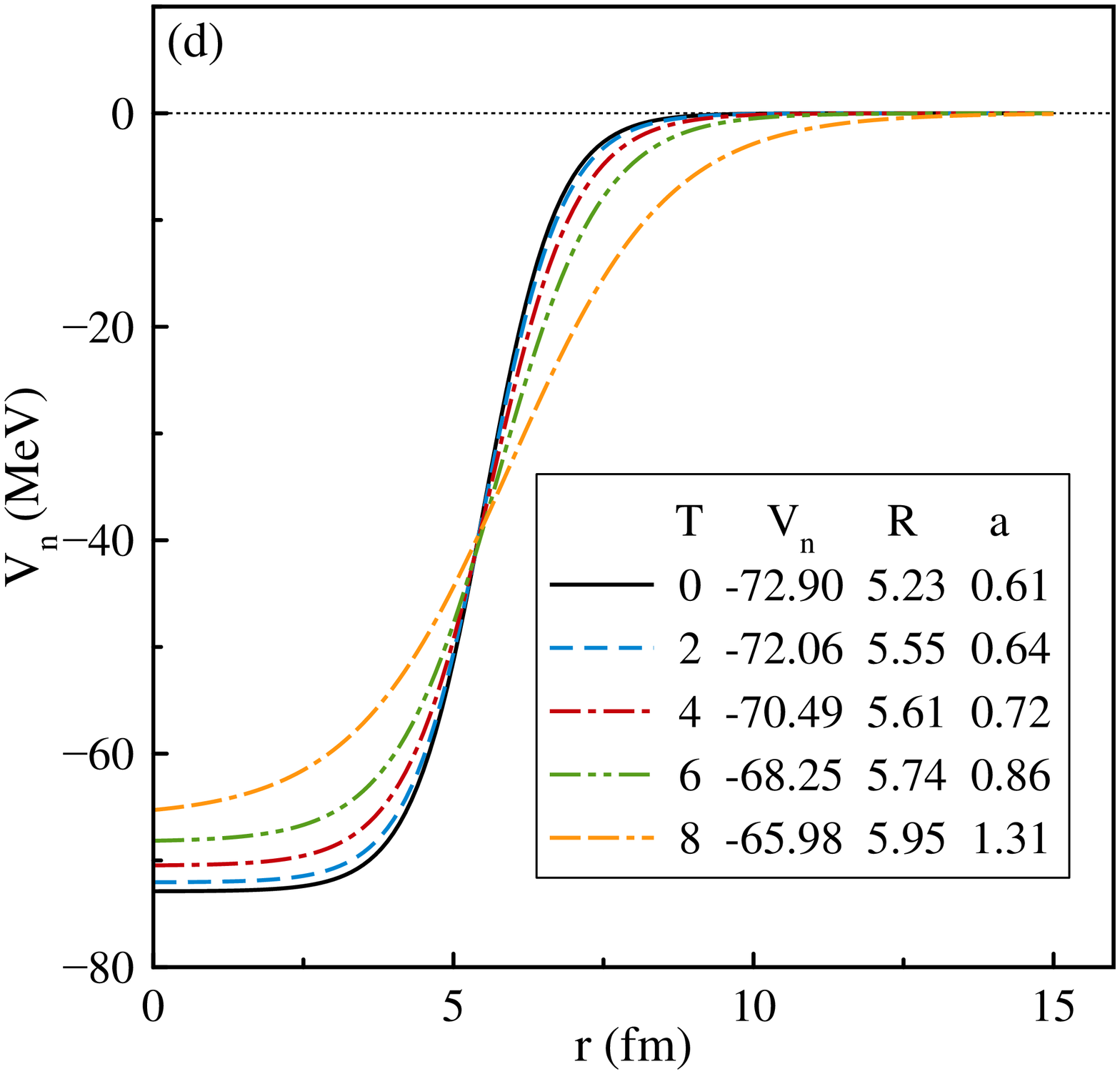}}\hfill
	\subfigure{\label{fig:VpxTSn100}
		\includegraphics[width=6.3cm,height=6.3cm]{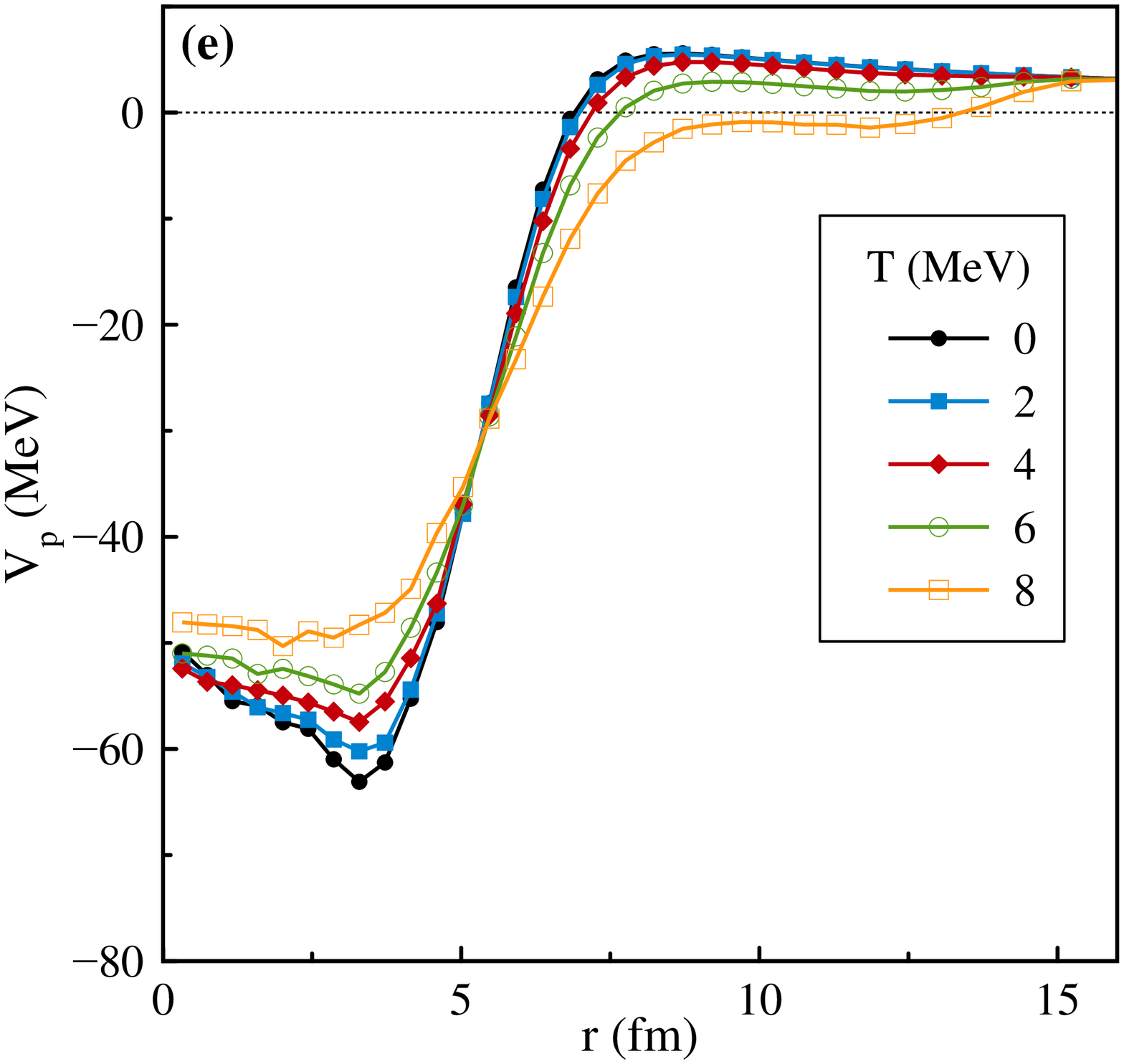}}\hfill
	\subfigure{\label{fig:VpxASn100FIT}
		\includegraphics[width=6.3cm,height=6.3cm]{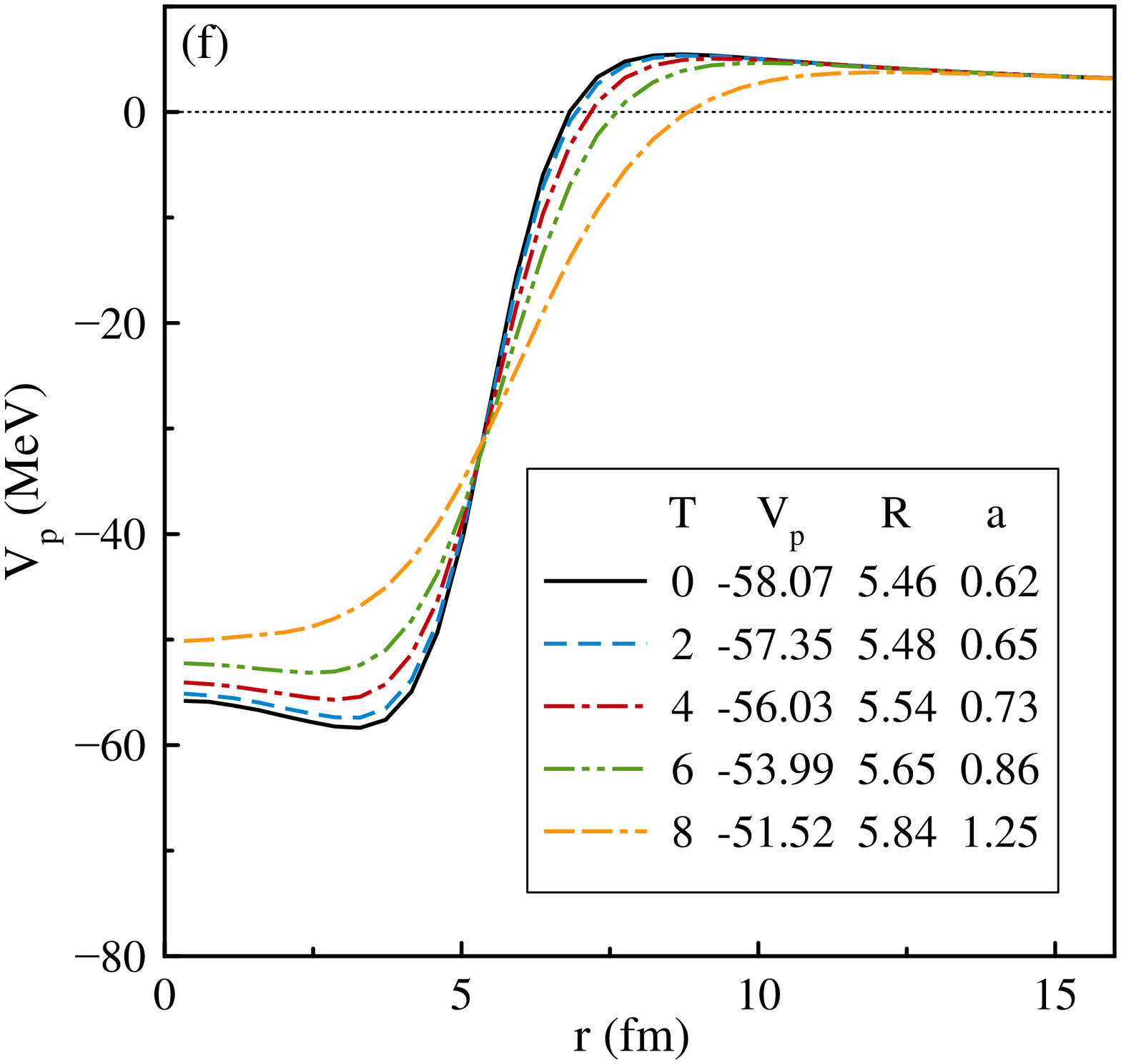}}\hfill
	\caption{\label{fig:VxTSn100} (Color online) Nuclear potentials of 
the nucleus $^{100}$Sn as a function of the radial distance with 
temperatures varying from $T=0$ to $T=8$ MeV. The left column represents 
our FTDHB calculations for (a) $\Sigma_c(r)$, (c) $V_n(r)$, 
and (e) $V_p(r)$. The right column shows our fit with a Woods-Saxon shape 
of the (b) $\Sigma_c(r)$, (d) $V_n(r)$, and (f) $V_p(r)$, together with 
the Woods-Saxon parameters depth $V_0$ (in MeV), radius $R$ (in fm), and 
surface diffuseness $a$ (in fm). \vspace{-.5cm}}
\end{figure*}

In order to study the same effect at finite temperature, we will 
fit the self-consistent potentials $\Sigma_c(r)$, $V_{n}(r)$, 
and $V_{p}(r)$ to a Woods-Saxon shape for $T\neq 0$.
In the left column of Fig.~\ref{fig:VxTSn100}, we show our FTDHB calculations
of the (a) $\Sigma_c(r)$, (c) $V_n(r)$, and (e) and $V_p(r)$ potentials as
a function of the radial distance for the nucleus $^{100}$Sn, 
in equilibrium with
the external gas, as the temperature varies from $T=0$ to $T=8$ MeV.
At $T=0$, the $\Sigma_c(r)$ [in Fig.~\ref{fig:VcxTSn100}] and $V_n(r)$
[in Fig.~\ref{fig:VnxTSn100}] potentials vanish at the
surface. When the temperature is increased, these 
potentials no longer go to zero
at large radii because of the contribution of the gas 
consisting of nucleons that
evaporate for $T\neq 0$. We show the potential $V_p(r)$ 
(in Fig.~\ref{fig:VpxTSn100}) for protons in $^{100}$Sn over
the same range of the temperatures. The proton potential vanishes at a larger
radius than the neutron one because of the long-range effect of
the Coulomb potential. In our calculations, we use the Bonche, Levit,
and Vautherin procedure to take into account the evaporated nucleons that
become important at temperatures above about $3 - 4$ MeV \cite{Bonche84,Bonche85}.
Note that beyond about $T\geq 8$ MeV, the nuclear structure is 
almost completely dissolved since the stability of a hot nucleus depends 
on maintaining the balance between surface and Coulomb contributions, 
as discussed in Refs.~\cite{Bonche84,Bonche85,Lisboa10,Lisboa16}.

To study the effect of temperature
on pseudospin symmetry and its possible causes, and in view of the 
systematics uncovered in Ref.~\cite{Lisboa03} referred to above, we 
study the change in the shape of the self-consistent mean fields with 
temperature, which appears as changes with temperature of the 
Woods-Saxon parameters of the fitted potentials, namely their 
depth ($V_0$), radius ($R$), and diffusivity ($a$). We show in the right 
column of Fig.~\ref{fig:VxTSn100}
our fit with a Woods-Saxon shape of the (b) $\Sigma_c(r)$, (d) $V_n(r)$, and
(f) $V_p(r)$ potentials of the nucleus $^{100}$Sn, together
with the values of the corresponding Wood-Saxon parameters, for
temperatures varying from $T=0$ to $T=8$ MeV. The fit is good for
temperatures $T\leq 8$ MeV and the Bonche, Levit, and Vautherin procedure
can be considered adequate for our calculations in this temperature range.
In the right column of Fig.~\ref{fig:VxTSn100}, we can read the 
Woods-Saxon parameters in the legend of each subfigure. We observe that, as 
the temperature grows, the depths of the
potentials decrease while their radii and surface diffuseness 
parameters increase. The depth of the potentials decreases  about
$\sim 10\%$ between $T=0$ and $T=8$ MeV as the radii increase in about the
same ratio of $\sim 10\%$. However, the surface diffuseness increases at least
$50\%$ or more over the same range of temperatures.
The same studies were performed for $^{132}$Sn 
and $^{150}$Sn with similar results.
\begin{figure*}[!ht]
	\subfigure{\label{fig:depthcxTSn}
		\includegraphics[width=6.5cm,height=6.5cm]{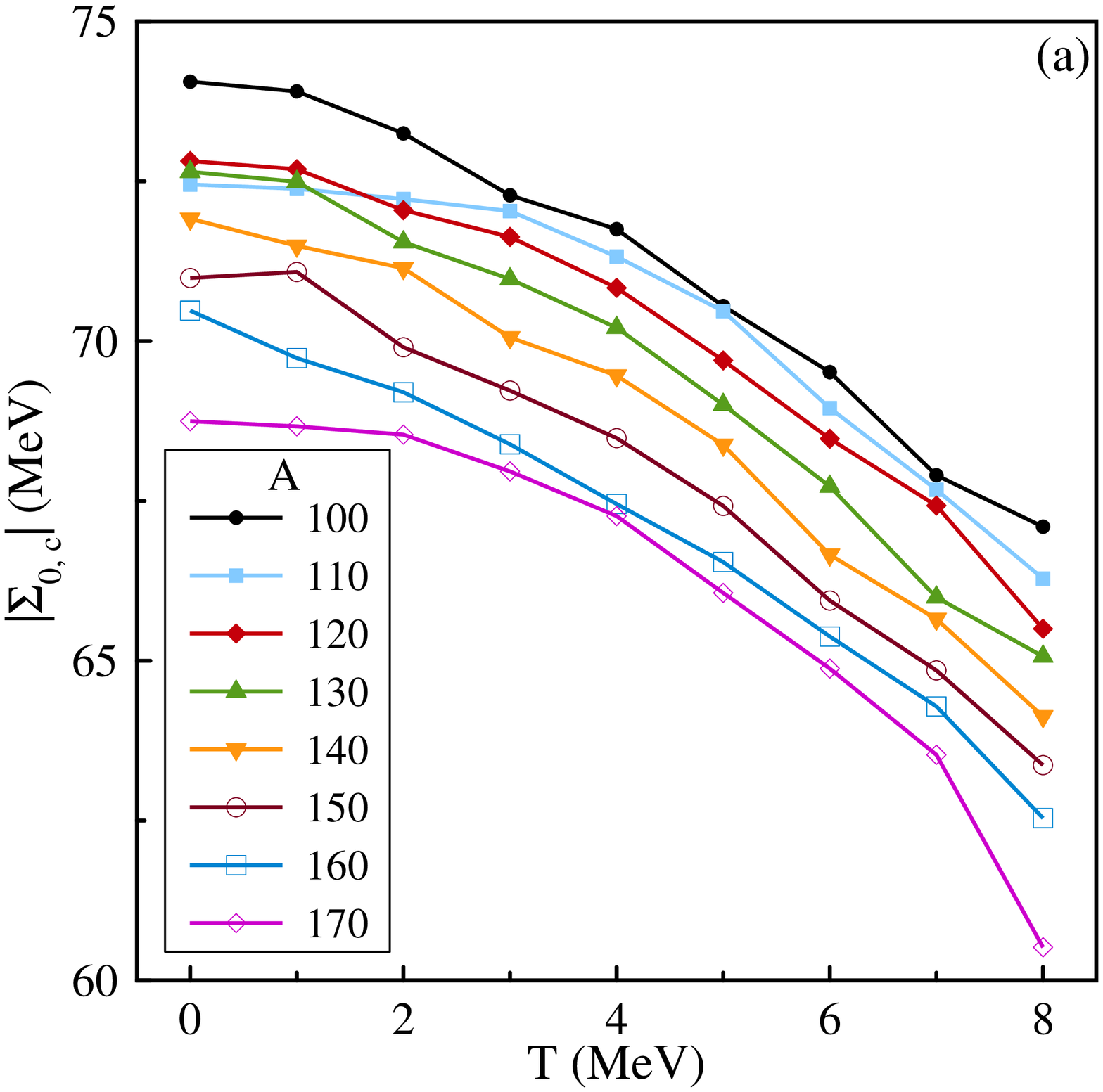}}\hfill
	\subfigure{\label{fig:radiicxTSn}
		\includegraphics[width=6.5cm,height=6.5cm]{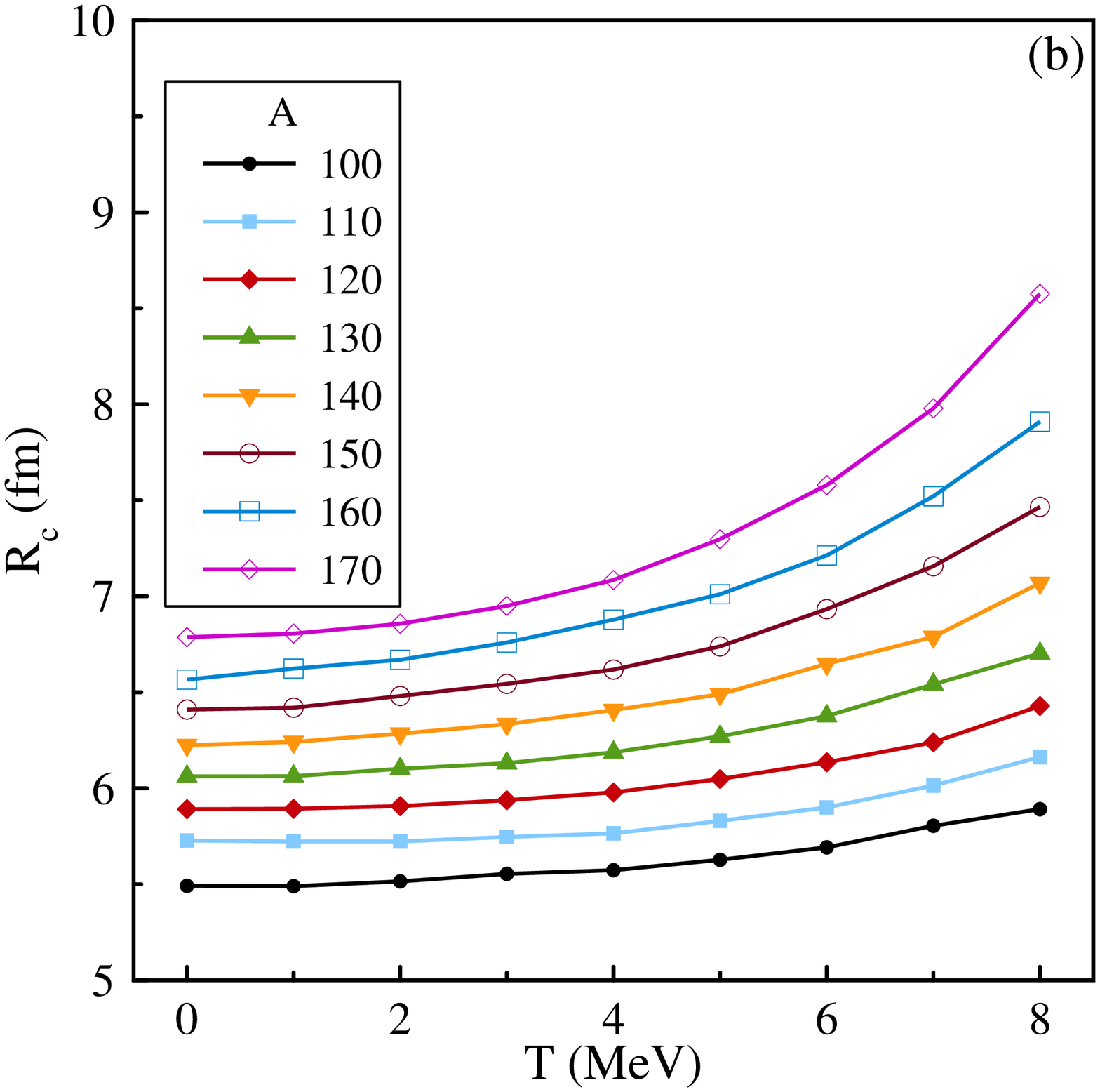}}\hfill\vspace*{-0.25cm}
	\subfigure{\label{fig:difuscxTSn}
		\includegraphics[width=6.5cm,height=6.5cm]{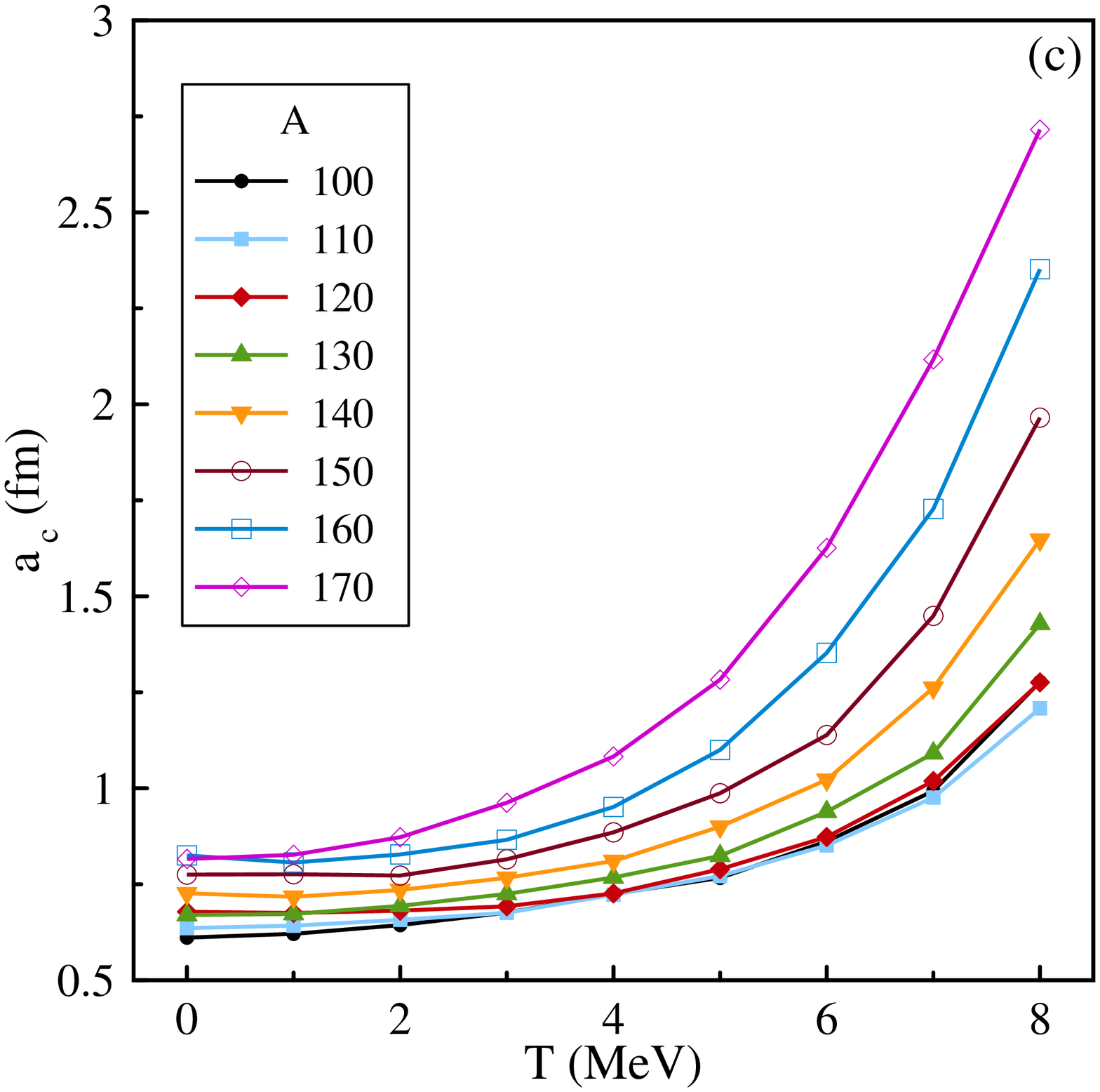}}
	\caption{\label{fig:SRaxT} (Color online)  The Woods-Saxon potential
    parameters for the (a) depth $|\Sigma_{0,c}|$ (in MeV), (b) radius $R_c$ (in fm),
    and (c) diffusivity $a_c$ (in fm), vs temperature for tin isotope chain.\vspace*{-0.35cm}}
\end{figure*}
%
\begin{figure*}[!ht]
	\subfigure{\label{fig:depthR2cxT}
		\includegraphics[width=6.5cm,height=6.5cm]{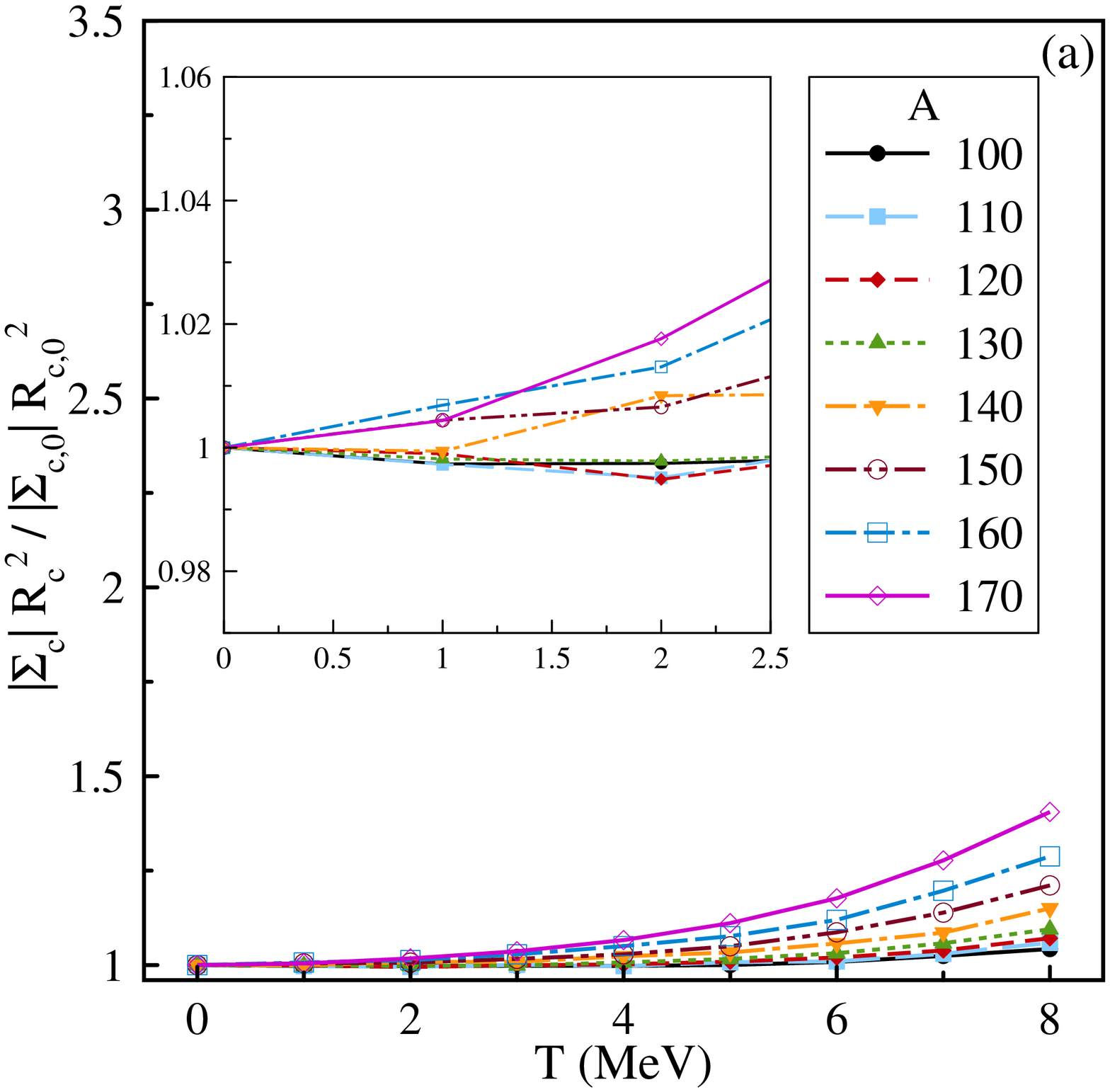}}\hfill
	\subfigure{\label{fig:acxT}
		\includegraphics[width=6.5cm,height=6.5cm]{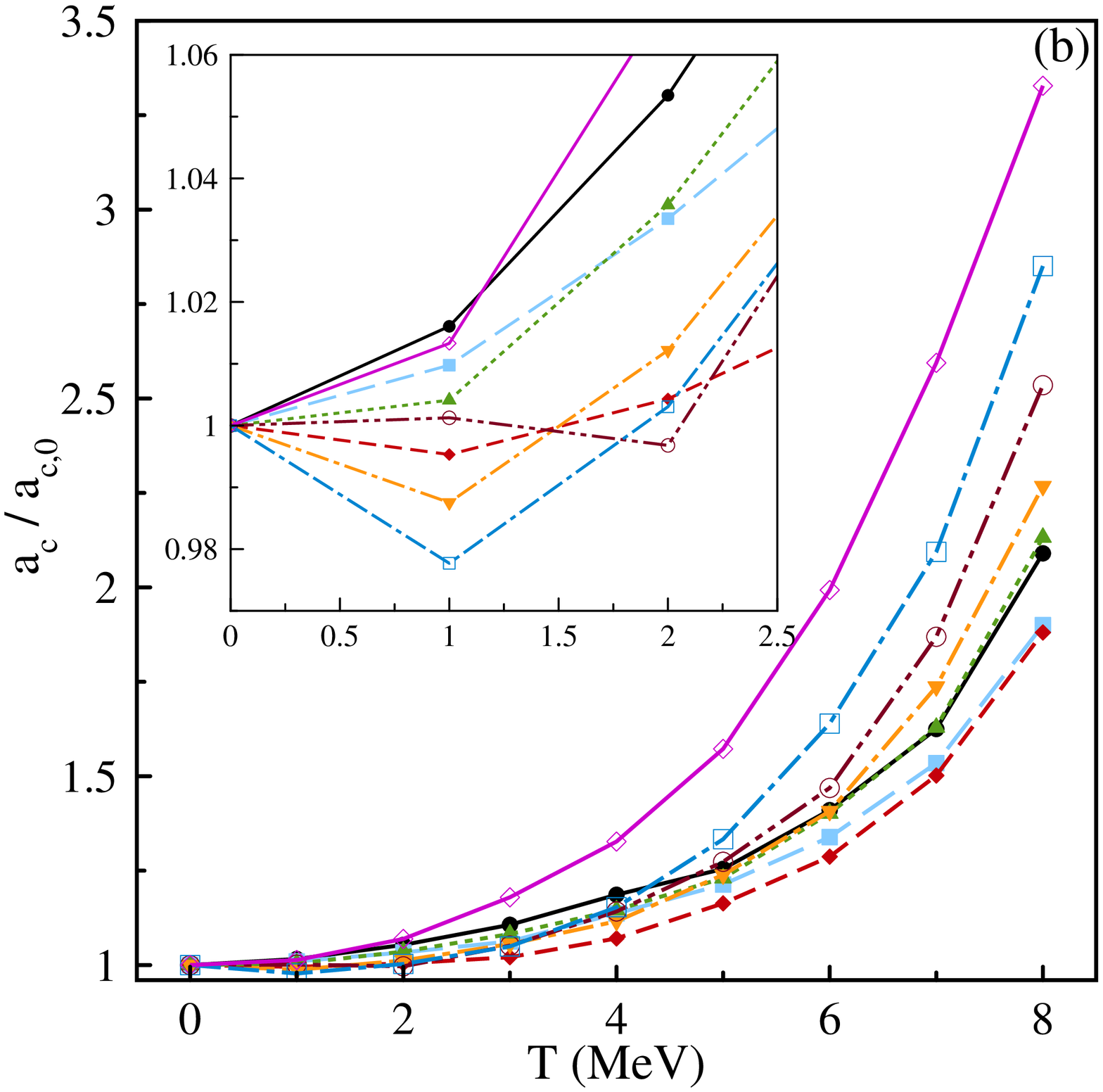}}\hfill
	\caption{\label{fig:ratioSRAxT} (Color online)  The Woods-Saxon parameters
		(a) $|\Sigma_{c}|R_c^2$ and  (b) diffusivity $a_c$
		divided by their respective values at $T=0$, vs temperature for the tin isotope chain.\vspace*{-0.35cm}}
\end{figure*}

One can conclude from these figures that when the temperature 
increases the central depth $\Sigma_0$ decreases and
the radius and the surface diffuseness increase. To see this more clearly,
in Fig.~\ref{fig:SRaxT} we show the Woods-Saxon parameters for the 
tin isotopes as the temperature grows up to the limit
$T=8$ MeV.
In Fig.\ref{fig:depthcxTSn}, the value of $|\Sigma_{0,c}|$ decreases
as the temperature increases.
In Fig.~\ref{fig:radiicxTSn}, we
show that the radius $R_c$ also increases with temperature.
However, as we see in Fig.~\ref{fig:difuscxTSn}, the surface diffuseness
increases quickly with increasing temperature. When we fix a value
of the temperature, the value of $|\Sigma_{0,c}|$ decreases as
$A$ increases, while both the radius $R_c$ and
surface diffuseness $a_c$ increase as $A$ increases. In fact, these results are
expected due to the known $A^{1/3}$ dependence of the nuclear radius.
Our results agree with calculations of the RMF 
theory at $T=0$ \cite{Alberto01,Alberto02}.

This opposing tendency of $|\Sigma_c|$ and $R$ produces values of
$|\Sigma_{c}|R_c^2$ that are roughly constant for each isotope from
$T=0$ to $T=8$ MeV. In Fig.~\ref{fig:ratioSRAxT}, 
we show the product $|\Sigma_{c}|R_c^2$
over $|\Sigma_{c,0}|R_{c,0}^2$ at $T=0$ for tin isotopes
as the temperature increases. The ratio is almost constant
and changes very little below $T=8$ MeV. In Fig.~\ref{fig:acxT}, we
show that diffuseness $a_c$ over $a_{c,0}$ at $T=0$ also increases 
with temperature, but this change is very large up to $T=8$ MeV in comparison
to that seen in Fig.~\ref{fig:depthR2cxT}.
Summarizing, as $T$ increases, the central depth $|\Sigma_0|$
decreases and the radius $R$ increases, but both effects 
balance each other, since
the values of $|\Sigma_0|R^2$ are roughly constant. Thus, when $T$ increases,
the dominant effect comes from the increasing diffuseness $a$, which favors the
pseudospin symmetry as found in Ref.~\cite{Lisboa03}.

In Table \ref{tab:psp100snN}, we show the pseudospin partners of
the neutrons and protons of $^{100}$Sn. The magnitude of the 
neutron pseudospin energy splitting
($\Delta E_n$ in MeV) decreases with increasing temperature.
We also see that the pseudospin splitting also decreases  
for protons ($\Delta E_p$ in MeV). However,
for protons, the pseudospin splitting is larger than for neutrons,
at least for the deep doublet $[2s_{1/2}-1d_{3/2}]$ of the symmetric
nucleus $^{100}$Sn.
%
\begin{table*}[!ht]
	\caption{Pseudospin energy splitting in MeV of the pseudospin partners of
     $^{100}$Sn for neutrons ($\Delta E_n$) and protons $(\Delta E_p)$
	 at several values of the temperature $T$ in MeV.}
	\begin{ruledtabular}
		\begin{tabular}{c|ccc|ccc||ccc|ccc|ccc}
			$T$
			& $2s_{1/2}$  & $1d_{3/2}$   & $\Delta E_n$
			& $2p_{3/2}$ & $1f_{5/2}$    & $\Delta E_n$
			& $2s_{1/2}$ & $1d_{3/2}$    & $\Delta E_p$
			& $2p_{3/2}$ & $1f_{5/2}$    & $\Delta E_p$ \\
			\hline
   0 & 34.38 & 38.74 & 4.36 & 20.94 & 24.23 & 3.29 & 19.25 & 23.74 & 4.49 &6.44 &9.64 & 3.20\\
 1.0 & 34.38 & 38.60 & 4.22 & 20.97 & 24.18 & 3.21 & 19.28 & 23.63 & 4.36 &6.49 &9.62 & 3.13\\
 2.0 & 34.24 & 37.80 & 3.56 & 21.01 & 23.81 & 2.79 & 19.32 & 23.03 & 3.71 &6.69 &9.45 & 2.75\\
 3.0 & 33.87 & 37.06 & 3.19 & 20.93 & 23.46 & 2.52 & 19.17 & 22.50 & 3.33 &6.82 &9.30 & 2.48\\
 4.0 & 33.45 & 36.38 & 2.93 & 20.87 & 23.14 & 2.27 & 18.95 & 22.01 & 3.06 &6.96 &9.18 & 2.22\\
 5.0 & 33.00 & 35.62 & 2.62 & 20.88 & 22.81 & 1.93 & 18.73 & 21.47 & 2.74 &7.19 &9.07 & 1.88\\
 6.0 & 32.43 & 34.70 & 2.27 & 20.90 & 22.45 & 1.55 & 18.47 & 20.83 & 2.37 &7.50 &9.00 & 1.50\\
 7.0 & 31.69 & 33.56 & 1.87 & 20.87 & 22.03 & 1.15 & 18.12 & 20.07 & 1.95 &7.85 &8.95 & 1.10\\
 8.0 & 30.74 & 32.10 & 1.36 & 20.76 & 21.42 & 0.67 & 17.68 & 19.10 & 1.42 &8.25 &8.85 & 0.60\\
		\end{tabular}
	\end{ruledtabular}
	\label{tab:psp100snN}
\end{table*}
%
\begin{table*}[!ht]
	\caption{Pseudospin energy splitting for neutrons ($\Delta E_n$) in MeV for pseudospin partners of $^{150}$Sn for several values of $T$ in MeV.}
	\begin{ruledtabular}
		\begin{tabular}{c|ccc|ccc|ccc|ccc}
	$T$
	& $2s_{1/2}$ & $1d_{3/2}$ & $\Delta E_n$
	& $2p_{3/2}$ & $1f_{5/2}$ & $\Delta E_n$
	& $2d_{5/2}$ & $1g_{7/2}$ & $\Delta E_n$
	& $2f_{7/2}$ & $1h_{9/2}$ & $\Delta E_n$  \\
	\hline
   0 & 32.91 & 35.78 & 2.87 & 21.93 & 24.66 & 2.73 & 11.79 & 13.29 & 1.50 & 3.11 & 2.52 & -0.59 \\
 1.0 & 32.83 & 35.73 & 2.90 & 21.87 & 24.67 & 2.79 & 11.77 & 13.34 & 1.57 & 3.12 & 2.59 & -0.53 \\
 2.0 & 32.62 & 35.57 & 2.94 & 21.79 & 24.71 & 2.91 & 11.85 & 13.54 & 1.69 & 3.32 & 2.90 & -0.42 \\
 3.0 & 32.31 & 35.20 & 2.89 & 21.74 & 24.57 & 2.83 & 12.06 & 13.66 & 1.59 & 3.73 & 3.27 & -0.46 \\
 4.0 & 31.87 & 34.63 & 2.75 & 21.69 & 24.31 & 2.62 & 12.37 & 13.77 & 1.40 & 4.31 & 3.77 & -0.55 \\
 5.0 & 31.29 & 33.81 & 2.52 & 21.60 & 23.91 & 2.31 & 12.73 & 13.87 & 1.14 & 5.04 & 4.37 & -0.67 \\
 6.0 & 30.53 & 32.73 & 2.20 & 21.45 & 23.37 & 1.92 & 13.13 & 13.95 & 0.82 & 5.91 & 5.07 & -0.83 \\
 7.0 & 29.51 & 31.32 & 1.81 & 21.17 & 22.64 & 1.47 & 13.54 & 14.00 & 0.47 & 6.93 & 5.91 & -1.02 \\
 8.0 & 28.17 & 29.48 & 1.31 & 20.72 & 21.66 & 0.94 & 13.95 & 13.97 & 0.03 & 8.16 & 6.87 & -1.30 \\
	\end{tabular}
	\end{ruledtabular}
	\label{tab:psp150SnN}
\end{table*}
%
\begin{table*}[!ht]
	\caption{Pseudospin energy splitting for protons ($\Delta E_p$) for the pseudospin partners of $^{150}$Sn for several values of $T$.}
	\begin{ruledtabular}
		\begin{tabular}{c|ccc|ccc|ccc|ccc}
			$T$
			& $2s_{1/2}$ & $1d_{3/2}$ & $\Delta E_p$
			& $2p_{3/2}$ & $1f_{5/2}$ & $\Delta E_p$
			& $2d_{5/2}$ & $1g_{7/2}$ & $\Delta E_p$
			& $2f_{7/2}$ & $1h_{9/2}$ & $\Delta E_p$  \\
			\hline
0  & 33.10 & 36.57 & 3.47 & 22.29  & 25.61 & 3.31 &  -      &  -       &  -      &  -     &  -     &  -     \\
1.0  & 33.13 & 36.60 & 3.47 & 22.32  & 25.69 & 3.37 & 11.87 & 14.35 & 2.48 &  -     &  -     &  - \\
2.0  & 33.14 & 36.51 & 3.37 & 22.39  & 25.85 & 3.46 & 12.02 & 14.70 & 2.69 &  2.25 & 3.69 &  1.43 \\
3.0  & 33.00 &36.25  & 3.24 & 22.46  & 25.81 & 3.35 & 12.32 & 14.92 & 2.60 &  2.80 & 4.16 &  1.37 \\
4.0  & 32.77 & 35.84 & 3.07 & 22.57  & 25.71 & 3.14 & 12.77 & 15.18 & 2.41 &  3.59 & 4.79 &  1.20 \\
5.0  & 32.45 & 35.29 & 2.83 & 22.71  & 25.55 & 2.84 & 13.35 & 15.49 & 2.13 &  4.61 & 5.58 &  0.97 \\
6.0  & 32.01 & 34.52 & 2.51 & 22.82  & 25.29 & 2.47 & 14.02 & 15.80 & 1.79 &  5.82 & 6.49 &  0.67 \\
7.0  & 31.32 & 33.44 & 2.12 & 22.82  & 24.84 & 2.03 & 14.70 & 16.07 & 1.37 &  7.20 & 7.50 &  0.30 \\
8.0  & 30.29 & 31.92 & 1.62 & 22.62  & 24.09 & 1.47 & 15.36 & 16.19 & 0.83 &  8.78 & 8.55 & -0.23 \\
		\end{tabular}
	\end{ruledtabular}
	\label{tab:psp150SnP}
\end{table*}

In Table \ref{tab:psp150SnN}, we show the neutron pseudospin partners 
of $^{150}$Sn.
The magnitude of the pseudospin splitting increases 
up to $T=2$ MeV and then begins to decrease with the temperature for the 
deeper doublets.
The splitting of the doublet $[2f_{7/2}-1h_{9/2}]$ has the opposite sign and 
its magnitude decreases up to $T=2$ MeV and then starts to increase with temperature.
In Table \ref{tab:psp150SnP} we show the proton pseudospin 
partners of $^{150}$Sn.
The magnitude of the pseudospin splitting of the
doublets $[2p_{3/2}-1f_{5/2}]$ and $[2d_{5/2}-1g_{7/2}]$ increases 
up to $T= 2$ MeV and then starts to decrease. The splitting of the 
doublets $[2s_{1/2}-1d_{3/2}]$
and $[2f_{7/2}-1h_{9/2}]$ decreases  with temperature already from $T=0$. 
The doublet
$[2d_{5/2}-1g_{7/2}]$ is not populated at temperature $T=0$ 
and  $[2f_{7/2}-1h_{9/2}]$
is populated only when $T\geq 2$ MeV. The non occupied states at $T=0$ are 
due to the temperature
effect when we consider the Fermi occupation factor. 
If we analyze each pseudospin
partner of Tables \ref{tab:psp150SnN} and \ref{tab:psp150SnP}, we see
that the energy splitting
is smaller for neutrons in comparison to protons, as expected from the 
isospin asymmetry of the pseudospin symmetry \cite{Alberto01, Lisboa03}.
\begin{figure*}[!ht]
	\subfigure{\label{fig:DExTSn150N} \includegraphics[width=7.5cm,height=7.5cm]{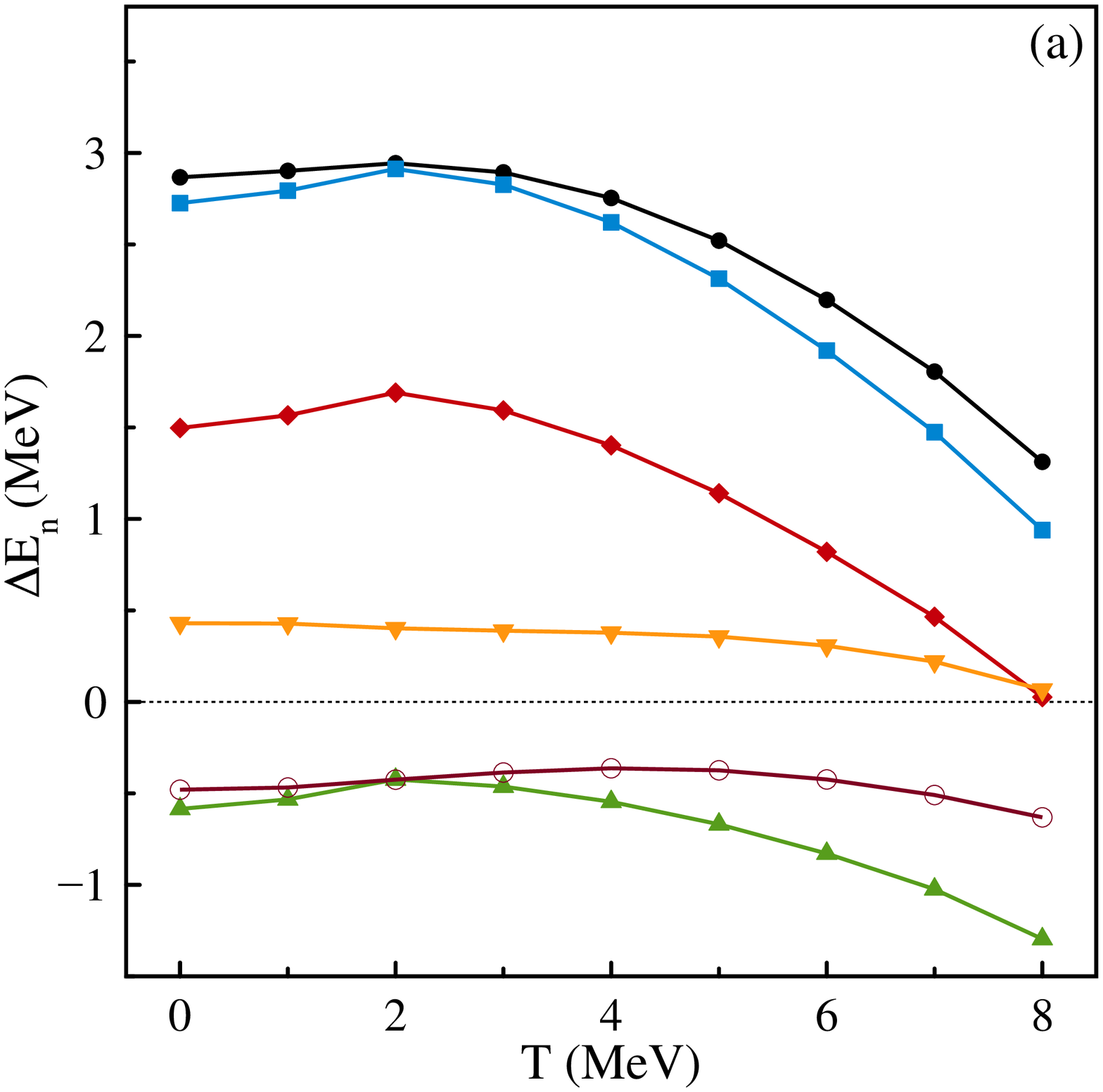}}\hfill
	\subfigure{\label{fig:DExTSn150P} \includegraphics[width=7.5cm,height=7.5cm]{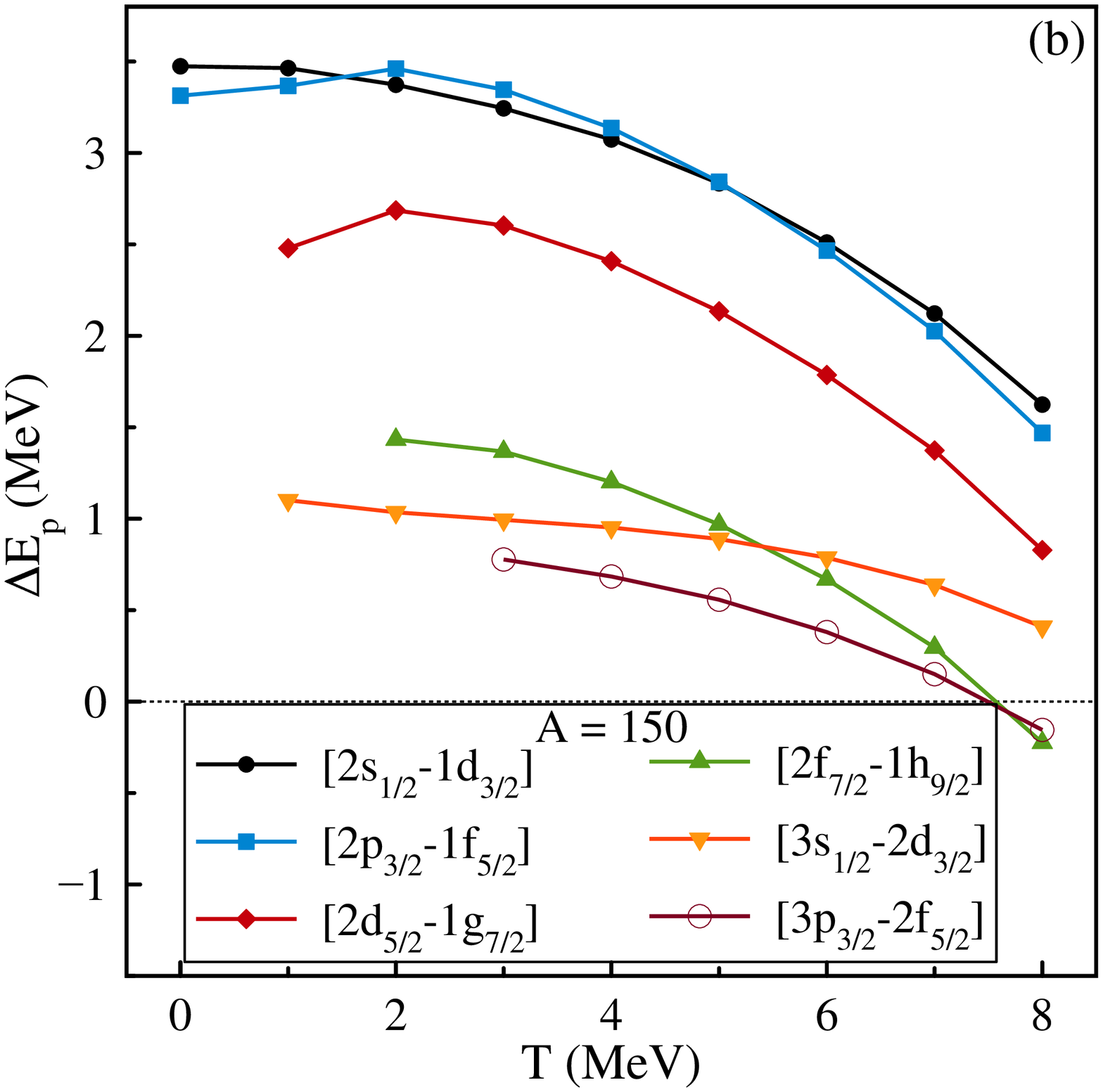}}\hfill
	\caption{\label{fig:partnes100SnT} (Color online)  Energy splitting for
		several pseudospin doublets of $^{150}$Sn for
		(a) neutrons
		and
		(b) protons for temperatures varying from $T=0$ up to $T=8$ MeV.}
\end{figure*}

 In Fig.~\ref{fig:DExTSn150N}, we show several neutron pseudospin doublets
of the nucleus $^{150}$Sn. 
The magnitude of the energy 
splitting increases  with temperature up 
to $T=2$ MeV and then begins to decrease, except for the 
doublet $[3s_{1/2}-2d_{3/2}]$, which decreases  monotonically. 
The energy splittings of the two doublets $[2f_{7/2}-1h_{9/2}]$ 
and $[3p_{3/2}-2f_{5/2}]$ decreases with temperature 
up to $T=2$ MeV.
In Fig.~\ref{fig:DExTSn150P}, we show several proton pseudospin doublets 
of the nucleus $^{150}$Sn. As we see there, the magnitude of the 
energy splitting also increases 
for the deeper levels up to $T=2$ MeV and unoccupied states exist below 
this temperature for the upper levels. 
This behavior below 
$T=2$ MeV is not consistently observed for other isotopes. 
For example, as we see in the zoom inside Figs.~\ref{fig:depthR2cxT} 
and \ref{fig:acxT}, respectively, the 
product $|\Sigma_{c,0}|\,R_{c,0}^2$ decreases  less than 1\% while 
the surface diffuseness $a_c$ increases  almost 5\% 
for $^{100}$Sn (black line). These effects could 
corroborate a decrease in the pseudospin 
splitting from $T=0$ MeV up for all levels of $^{100}$Sn, as we see 
in Table \ref{tab:psp100snN}.
However, these variations are in general small and sometimes 
irregular for the temperatures  below $T = 2$ MeV
and thus do not allow one to establish a clear dependence 
between the parameters of the Wood-Saxon 
potential and the pseudospin splittings.
This is not the case for $T>2$ MeV, because the surface diffuseness increases more
than product $|\Sigma_{c,0}|\,R_{c,0}^2$ as the temperature increases, 
and the energy splitting of the pseudospin doublets becomes small. The exceptions 
are the energy splittings of the two
doublets $[2f_{7/2}-1h_{9/2}]$ and $[3p_{3/2}-2f_{5/2}]$ for neutrons.
The former becomes less degenerate above $T=2$ MeV, while 
the latter becomes less degenerate above $T=4$ MeV, as we see
in Fig.~\ref{fig:DExTSn150N}. 

The increase of diffusivity with temperature determines 
the growth of the ratio $a_c/a_{c,0}$, which is larger than the 
changes induced by the temperature in 
$|\Sigma_{c}|R_c^2/|\Sigma_{c,0}|R_{c,0}^2$, favoring the pseudospin symmetry.

\section{Conclusion}
\label{sec:conclusion}

	In this work we have studied the effects of temperature on the 
energy splitting of several pseudospin doublets of the spherical tin 
isotopes. We used the finite temperature Dirac-Hartree-Bogoliubov 
(FTDHB) formalism to obtain the mean-field and Coulomb potentials in 
a self-consistently 
calculation \cite{Lisboa16}. This
formalism allows us to take into account the pairing and deformation beyond
the mean field and Coulomb potentials. 
In our calculations, we observe that the mean field potentials have the shape 
of Woods-Saxon potentials
for the temperature range from $T=0$ to $T=8$ MeV. By fitting the potentials 
to Wood-Saxon potentials, we were able to investigate the correlation 
between the 
pseudospin splittings and the parameters of the Wood-Saxon potential: the 
depth ($\Sigma_0$), 
surface diffuseness ($a$), and radius $(R)$.
We studied the tin nuclei from $A=100$ to $A=170$ as a function of 
temperature between
$T=0$ and $T=8$ MeV. For each nuclei we obtained the values of the parameters 
of the Wood-Saxon potential and analyzed their variation with 
increasing temperature. 
We found that
for $^{100}$Sn the depth of the potential decreases  while the radius and
surface diffuseness increases  with temperature. The depth of the 
potential decreases  
on the order of $\sim 10\%$ while 
the radius increases  in the same ratio between $T=0$ and $T=8$ MeV. However, the
diffusivity increases  by at least $\sim 50\%$ in the same temperature range.
The other tin isotopes show similar results. 
From the calculation of the energy splittings for the neutron and proton 
pseudospin partners of the tin isotopes at several values of the temperature, 
we see that in general the pseudospin energy splittings 
decrease with temperature.
This confirms the systematics already found in Ref. \cite{Lisboa03} for the 
tin isotopes at $T=0$, in which the change in diffusivity was the main 
driver for 
the variation in pseudospin energy splittings, which favors 
pseudospin symmetry. The decrease of the energy difference 
between pseudopsin doublets with the increase of the temperature seems
also to be valid for deformed nuclei at large temperatures. 
In Ref.~\cite{Lisboa16}, 
we show that $^{168}$Er becomes spherical and the splitting decreases  
for temperature $T\geq 4$ MeV.

We can thus restate, now including the effects of temperature that, in general, 
there is a correlation between the shape of the nuclear mean-fields, described 
here by Wood-Saxon parameters, and the onset of pseudospin symmetry on nuclei.
\begin{acknowledgments}
The authors acknowledge financial support from CAPES, CNPq, FAPESP, and FAPERN. R.L. acknowledges, in particular, support from the FAPERN (Foundation for Research Support of Rio Grande do Norte, Project No. 003/2011).  M.M. acknowledges the financial support of CNPq and FAPESP (Foundation for Research Support of the State of São Paulo, Thematic Project No. 2013/26258-4). B.V.C. acknowledges support from the CNPq (Project No. 306692/2013-9) and from FAPESP  (Thematic Project No. 2017/05660-0). P.A. would like to thank CFisUC for travel support. This work is a part of the project INCT-FNA (National Institutes of Science and Technology - Nuclear Physics and Applications, Proc. No. 464898/2014-5).
\end{acknowledgments}

\end{document}